\documentclass[usenatbib,twocolumn]{art}
\usepackage{graphicx}
\usepackage{epstopdf}
\usepackage{subfloat}
\usepackage{subfigure}
\usepackage{hyperref}
\usepackage{multirow}
\pdfoutput=1
\def\araa{ARAA}
\def\mnras{MNRAS}
\def\apj{APJ}
\def\aj{AJ}

\def\aap{AAP}
\def\aaps{AAPS}

\def\nat{Nature}
\def\na{NewAstro}

\def\apss{APSS}

\def\HI{{H~{\sc i} }}
\begin{document}

\title{Estimating statistics of sky brightness using radio interferometric observations}
\author[Prasun Dutta and Meera Nandakumar]
{Prasun Dutta$^{1}$\thanks{Email:pdutta.phy@itbhu.ac.in},  
Meera Nandakumar$^{1}$\thanks{Email: meeranandakr.rs.phy17@itbhu.ac.in}, 
\\$^{1}$ Department of Physics, IIT (BHU) Varanasi, 221005  India. 
}
\maketitle 

\begin{abstract}
Radio interferometric data are used to estimate the sky brightness distributions in radio frequencies. Here we focus on  estimators of the large-scale structure  and the power spectrum of the sky brightness distribution inferred from  radio  interferometric observations and assess their efficacy using simulated observations of the model sky. We find that while the large-scale distribution can be unbiasedly estimated from the reconstructed image from the interferometric data, estimates of  the power spectrum of the intensity fluctuations  calculated from the  image are generally  biased. The bias is  more pronounced for diffuse emission. The visibility based power spectrum estimator, however,  gives an unbiased estimate of the true power spectrum.  We conclude that for an observation with  diffuse emission the reconstructed image can be used to estimate the large-scale distribution of the intensity, while to estimate the power spectrum, visibility based methods should be preferred.
\end{abstract}


\section{Introduction}
Radio interferometers are the key instruments to map spatially resolved sky brightness fluctuations at the radio frequencies \citep{1999ASPC..180.....T}. Within radio frequencies, there exist different emission mechanisms \citep{1985rpa..book.....R}, continuum emission by synchrotron radiation, 21-cm line emission from the neutral hydrogen gas or \HI are  to name a few.  Various statistics of the  sky brightness distribution are evaluated from the radio interferometric data  to  investigate different properties of the galactic and extragalactic sources including  the magnetic field as well as relativistic electron distributions in  galaxies \citep{1989Natur.340..537S, 2013MNRAS.433.1675B}, morphology and dynamics of \HI in the nearby dwarf and spiral galaxies \citep{2008MNRAS.386.1667B, 2008AJ....136.2563W, 2010MNRAS.405L.102D, 2009MNRAS.398..887D, 2013NewA...19...89D}, origin and evolution of  radio jets and lobes in  radio galaxies \citep{2012BASI...40..121N}, structures of the supernovae remnants \citep{2009MNRAS.393L..26R}, the cosmological evolution of \HI \citep{2011MNRAS.411.2426G} etc. 

A class of these investigations uses large-scale distribution of the source  emission  to infer about local physical properties like morphology of supernovae remnants, radio galaxies and spiral galaxies, radial distribution of \HI  and rotational velocities or relation between the star formation and the gas distribution in the spiral galaxies etc. from the one point statistics like mean or median of the specific intensity at different positions within the source. On the other hand, two-point statistics, like the autocorrelation function, the structure function or the power spectrum of the sky brightness fluctuations carry several important physical informations like the  properties of MagnetoHyDrodynamic (MHD) turbulence in the supernovae remnants  or hydrodynamic turbulence in the \HI in the galaxies, evolution of the distribution of \HI during the cosmic dawn, epoch of reionization and post reionization era etc. In most of these cases, a power spectrum of the specific intensity is evaluated from the observed interferometric data \citep{2013MNRAS.436L..49D, 2012ApJ...754...29Z}. 

Radio interferometers inherently measure a quantity called visibility, a complex transform of the sky brightness distribution. Roughly speaking, for many of the science cases described above, the visibilities can be approximated as the Fourier transform of the sky brightness distribution measured at certain spatial frequencies. Estimation of the one point statistics of the sky brightness distribution requires a faithful reconstruction of it (the reconstruction is usually termed as the image)  from the visibilities. As the visibilities are often not measured at all the spatial frequencies, the reconstruction of the image is not straightforward and involves complicated algorithms to deconvolve the effect of this incomplete measurement. 

One approach to estimate the two-point statistics is to  use the reconstructed image and calculate either the structure function or the autocorrelation function in the image plane or estimate the power spectrum of the image in its Fourier conjugate plane. We shall call the estimators  of two point statistics, that rely on the reconstructed image,  the {\em image based estimators} (e.g \citet{2012ApJ...754...29Z, 2014MNRAS.441..525W,2017MNRAS.466.1093G}). Using the image based estimators it is possible to evaluate the two-point statistics of a part of the astrophysical source in consideration.  This is essential in some particular cases, viz., in the  correlation of star formation with the turbulence in the interstellar medium (ISM), a variation of the MHD turbulence in the arm and inter arm regions of the spiral galaxies \citep{2013MNRAS.433.1675B} etc. However, any problem in the image reconstruction is likely to show up as artefacts in the image based estimation of the two-point statistics and need to be investigated thoroughly before using for scientific inference.  On the other hand, a different class of estimators of the two-point statistics are used in  literature where visibilities are directly used to estimate the power spectrum of the sky brightness distribution  \citet{2009MNRAS.398..887D, 2016MNRAS.463.4093C}. As this does not need the image reconstruction, these estimators are more direct and not prone to artefacts of image reconstruction. We shall call these estimators the {\em visibility based estimators}. As the visibility based estimators use the visibilities directly, they can not be used to estimate the two-point statistics of a part of the image. However, for a few particular cases, these limitations can be overcome using suitable  techniques (see e.g, \citet{2010MNRAS.405L.102D}).

Image reconstruction from the visibilities is a long-standing problem \citep{2017isra.book.....T}, we shall discuss it in detail in the next section. The objective here is  to have the best guess of the sky brightness distribution from the limited observations that an interferometer provides. The question is  how accurately the one and two-point statistics can be computed from these reconstructions. As the reconstruction process requires various input parameters from the astronomers,  the outcome is not unique (see e.g, \citet{1999ASPC..180..151C, 1998AJ....115.1693C, 1995ApJ...450..559B, 2007ApJ...654...99W, 1999ASPC..180.....T}). In this paper, we investigate the efficacy of different estimators of the first and second order statistics of the radio interferometric data using simulated observations.  Engaging with all different classes of radio interferometric observations as well as all different imaging technique is not possible within the scope of  a single paper. Here we have focused on a particular problem of estimating the first and second order statistics of the \HI emission of spiral galaxies. Our results are directly applicable, but not limited  to the cases involving the same class of problems.  

The rest of the paper is divided  as follows: Section (2) describes \HI observations and reconstruction of the image,  simulation of the interferometric observation is discussed in section (3), section (4) discusses  the visibility and image based statistical estimators, we present the result and analysis of the simulated data in section (5)  and conclude in section (6).

\section{Brief overview of Radio interferometric observation}
Here we give a brief overview of the  interferometric observation; the interested readers may refer to \citet{1999ASPC..180.....T} for further details.  Radio interferometers  are collection of many  array elements, called antennas, arranged in a specific pattern on the surface of the earth \footnote{In principle, there can be ``zero spacing'' interferometers \citep{2014arXiv1406.2585M}, interferometers with a very small number of baselines and the antenna can be kept in space. However, most of the interferometers used today consist of many antennae on the surface of the earth.}. Each antenna records the electric field   of the electromagnetic wave coming from a particular direction of the sky.  The physical size of the individual antenna and the observing wavelengths limits the sensitivity of the antenna as well as the entire interferometer to a limited portion of the sky. This is known as the field of view (FOV) of the interferometer. Electric fields from each antenna pair are correlated and recorded. This quantity is known  as the visibility function. The pair of antenna for which the visibility is recorded is called a baseline. The baseline vector $\vec{U}$ is the ratio of instantaneous projected separation of the antenna pair on a plane perpendicular to the direction of the incoming wave from the sky to the observing wavelengths. Clearly, the visibilities are functions of the baseline vector, i.e, $V(\vec{U})$. However, they can be measured only at the discrete values  $\vec{U_{i}}$, which correspond to the physical baselines offered by all  the  antenna pairs. We introduce a function $S(\vec{U})$ to capture this sampling of the baselines by the interferometer:
\begin{equation}
S(\vec{U})=\sum^{N_{b}}_{i=1}\delta_{D}(\vec{U}-\vec{U_{i}}).
\end{equation}
Henceforth, our discussion will be limited to the observation of \HI 21-cm radiation from nearby spiral galaxies. Typically, these galaxies are about $\sim 2 -20$ Mpc away and hence their extent in the sky is limited to $\sim 40'$ or lower.  Interferometers with relatively narrow FOV of about one degree is sufficient to observe these galaxies. The existing radio interferometers like VLA \footnote{NRAO-VLA: Very Large Array, New Mexico}, GMRT \footnote{Giant Meterwave Radio Telescope, NCRA-TIFR} etc.,  have the required FOV.  In such cases, the baseline vector can be assumed  to be two dimensional and the visibilities can be approximated as the Fourier transform of the sky brightness distribution \citep{1999ASPC..180..419S} sampled at the baselines where the interferometric measurements are done, i.e,
\begin{equation}
V(\vec{U})=\tilde{I}(\vec{U})S(\vec{U}) + \mathcal{N}(\vec{U}),
\label{eqn:visdef}
\end{equation} 
were $\tilde{I}(\vec{U})$ is the Fourier transform of the sky brightness distribution $I(\vec{\theta})$ and  $\mathcal{N}(\vec{U})$ is the measurement noise. This is the quantity directly  measured by the radio interferometers. Note that, to completely describe a real measurement, we need to also scale the first term on the  right hand side of the eqn~(\ref{eqn:visdef})  by the gain of the interferometer.  However, here we assume that  a proper calibration procedure is followed to take care of the effect of the gain. Note that, we have neglected the effect of the antenna primary beam here. This is justified if the angular extent of the galaxy is much smaller than the full width of half maxima ( FWHM) of the primary beam of the antenna. On the other hand, if the angular extent of the galaxy is larger, the  primary beam can be included in the window function as described in eqn~(\ref{eq:mod}) in the next section.

The visibilities, as given by eqn~(\ref{eqn:visdef}), are used directly in the visibility based estimators of the two-point statistics of the sky brightness distribution. However, to estimate the one-point statistics at different points in the sky or to estimate the two-point statistics using an image based estimator, reconstruction of the sky brightness distribution is necessary from the observed visibilities. 

Inverse Fourier transform of the measured visibility is called the dirty image:
\begin{equation}
 I_{D}(\vec{\theta})=I(\vec{\theta}) \otimes B_{D}(\vec{\theta}).
\label{eq:dbeam}
\end{equation}
Here $B_{D}(\vec{\theta})$ is the Inverse Fourier transform of the weighted sampling function $S(\vec{U})$ and essentially the  Point Spread Function (PSF) of the interferometer.  The weighting schemes are discussed shortly. The symbol $\otimes$ denotes convolution here. We have neglected the measurement noise  for simplicity.  The PSF of the interferometer is often called the dirty beam as it has secondary maxima  around the centre, known as the side-lobes. Reconstruction of the sky brightness distribution is essentially  a deconvolution of the interferometer PSF from the dirty image. Since the sampling function can be quite discrete, and often irregular and incomplete, the interferometer PSF can be quite complicated thereby making the  deconvolution procedure  non trivial.  Different algorithms have been devised for this purpose  including types of  CLEAN \citep{1974AAS...15..417H, 1979AJ.....84.1122C, 1980AA....89..377C, 1984iimp.conf..333S}, Maximum Entropy Image Reconstruction  (MEM) \citep{1986ARAA..24..127N} and  RESOLVE \citep{2016AA...586A..76J}. In this paper, we focus on the CLEAN algorithm, which is the most widely used algorithm in the radio astronomy community to date \citep{1999ASPC..180..419S}.

Ever since the first version of CLEAN was outlined in \citet{1974AAS...15..417H} it has been widely used and also widely evolved. In CLEAN, the sky image is assumed to be a collection of point sources. The algorithm relies on estimating the  brightness and position of all the point sources in the sky from the image using an iterative procedure. This is achieved in different ways in different variations of CLEAN. Here we have used the Cotton-Schwab variant of  CLEAN \citep{1980AA....89..377C}.  The first step in any CLEAN algorithm is to make the dirty image from the observed visibilities. As the number of measured visibilities are rather large, Fast Fourier Transform (FFT) is used to make the dirty image. For most of the interferometers, the visibilities are not measured at regular intervals (like in a regular grid) but rather at apparently random locations  in the baseline plane. Hence, values of the visibility function are estimated through interpolations at regular intervals in the baseline plane to initiate FFT. This process is called gridding. Further, different grid points in the baseline plane usually have a different  number of measurements of the visibilities and they can be given different weights while constructing the image. This process is called weighting. Note that, some of these grid points may not have any measurements, in fact, a part of the baseline plane may be completely void of measured visibilities. This is termed as the incomplete baseline coverage of the interferometer. The weighting schemes try to partially rectify the effect of the incomplete coverage. Two extreme weighting schemes used are called uniform and natural weightings respectively. In the case of natural weighting, all the measurements are given the same weight and added together. This type of weighting emphasizes the part of the visibility plane where more measurements are present. Note that, when a weighting scheme is used, the effective dirty beam is to be considered as the Fourier transform of the sampling function multiplied by the weighting kernel. As for most of the interferometers, the baseline coverage is better at lower values of $\mid \vec{U}\mid$, the PSF derived from the naturally weighted visibilities are broader but with fewer side lobes. On the other hand, in a different weighting scheme called uniform weighting, the visibilities are weighted by the local density of measured visibilities before gridding. This type of weighting results in a narrower PSF but have a higher power in the  side lobes. The robust weighting scheme designed by \citet{1995AAS...18711202B}  tries to combine the natural and uniform weightings. 

Since the astronomical sources are usually confined to a part of the sky, in  principle the visibilities need to be measured in all parts of the infinite baseline plane. However, for a practical interferometer, there exists a longest baseline value $\mid \vec{U}_{max}\mid$, beyond which no measurements exist. In fact, the baseline coverage abruptly decreases at a certain baseline value for a given array configuration and observation conditions (like the declination of the source, the integration time etc.). This results in oscillations in the image plane (dirty as well as reconstructed image). To reduce this effect, the visibilities are multiplied by a function $T(\vec{U})$ of the baseline, which gradually falls to zero at the largest baseline of the interferometer. This works as a global weight in the visibility plane and  is called tapering. Usually, a Gaussian function is used where the user can specify the full width at half maxima of the Gaussian guessing it  from the sampling function. The detailed process of gridding and different weightings and their effects are discussed in \citet{2017isra.book.....T}.  We have used the default gridding convolution function in the task IMAGR in AIPS \footnote{NRAO AIPS: Astrophysical Image Processing System} which uses a spherical function. Note that the weighting  schemes are used to reduce the artefacts in the image that may arise from the incomplete baseline coverage. However, they may introduce additional effects, which one needs to assess and quantify before using the reconstructed image for scientific inferences.

Once the dirty image and the corresponding dirty beam is calculated from the observed visibilities, the Cotton-Schwab variant of the CLEAN progresses in the following way. The dirty beam is normalized to have unity value for the brightest pixel. The algorithm identifies the pixels in the dirty image that have the fractional intensity greater than the highest exterior side-lobes in the dirty beam. It then records the brightness and position of these pixels as CLEAN components. This serves as a temporary point source model of the sky. A  part of the discrete representation of the dirty beam is then multiplied by the above point source model and a fraction of it is subtracted from the image. This fraction is called the loop gain. The procedure repeats till it collects all the pixels having larger values of brightness than the brightest side-lobes in the dirty beam. This cycle is termed as a minor cycle. At the end of the minor cycle, the point source model is used to produce model visibilities at the location of the interferometer measurements in the baseline plane and is subtracted from the original measurements. This part, including the minor cycle it preceded,  is referred to as a major cycle. The process repeats itself through several major cycles and eventually generates a final point source model for the sky and a residual image.  A CLEAN or restoring beam is estimated by fitting the central part of the dirty beam with a Gaussian (though other functions can also be used, Gaussian is the most commonly used one). The point source model is then convolved with the restoring beam and the residual image is added. The latter provides a diagnostic  of the image reconstruction.  

The efficacy of CLEAN reconstruction algorithm, its limitations and known artefacts are discussed in detail in \citet{2017isra.book.....T}, we highlight a few relevant points here. It has been shown in \citet{1978AA....65..345S} and \citet{1979ASSL...76..261S} that in absence of noise and in the case when a lower number of point sources are required to construct the sky brightness distribution than the independent measurement of visibilities, CLEAN reduces to a least square image estimation procedure. However, in a real scenario, particularly with extended sources and measurement noise, CLEAN does not produce the unique reconstruction of the sky. The reconstructed image depends on many user selectable parameters, like the loop gain, size of the dirty beam patch used in the minor cycle, the weighting and tapering schemes etc. For example, it is recommended to use a smaller loop gain and a larger patch of the dirty beam in the  minor cycles in the case of diffuse emissions \citep{1999ASPC..180.....T}.

The uncorrelated measurement noise in the visibilities gives rise to correlated noise in any reconstructed image and can not be avoided. This is also the source of correlated noise present in image based estimates of  the power spectrum in case of incomplete baseline coverage (see   \citet{2015AA...581A..59J}).

\section{Simulating model visibility dataset}
In this article, we are interested in investigating the efficacy of different statistical estimators used to interpret the radio interferometric data. We proceed  by simulating an observation with a known sky model based on the  observed  \HI emission from the nearby external spiral galaxies (similar models are used in \citet{2009MNRAS.398..887D}). We write the  specific intensity distribution from such a galaxy as a function of the angle $\vec{\theta}$ (for simplicity the $x$ and $y$ component of the vector $\vec{\theta}$ can be considered along the local directions of Right Accession and Declination) from the centre of the galaxy as 
\begin{equation}
I(\vec{\theta})=W(\vec{\theta})\left[ \bar{I} + \delta I(\vec{\theta})  \right],
\label{eq:mod}
\end{equation}
where $W(\vec{\theta})$ quantifies the large-scale distribution  of the \HI column density and is normalized as $\int W(\vec{\theta}) d\vec{\theta} = 1$.  We call this the window function.  The quantity $ \bar{I}$ is the total intensity coming from the entire galaxy and $\delta I(\vec{\theta})$  corresponds to random fluctuations in the specific intensity. In the case of the ISM of spiral galaxies, such fluctuations arise as a result of compressible fluid turbulence therein. The mean of the fluctuations $\delta I(\vec{\theta})$  is zero. At a given $\vec{\theta}_{0}$ in the galaxy the ensemble  average of the term $W(\vec{\theta})  \delta I(\vec{\theta}) $ can be written as $W(\vec{\theta}_{0})  \langle \delta I(\vec{\theta}) \rangle$, where  $W(\vec{\theta}_{0})$ is the value at the window function at that $\vec{\theta}_{0}$. The window function falls to zero at large $\vec{\theta}$, which tapers  the random fluctuations $\delta I(\vec{\theta})$  at large $\vec{\theta}$ eventually ensuring zero specific intensity arising from the sky beyond the galactic extent. 
\footnote {Since we are mostly interested to estimate the power spectrum of  $\delta I(\vec{\theta})$, in general, the window function can be thought of as a multiplication of the galaxy window function with the primary beam of the interferometer. However, most of the cases the angular extent of the galaxy is smaller than the primary beam and the latter can be ignored.}
\subsection{Modeling window function}
The \HI profile of a spiral galaxy is dominated by the radial variation in \HI column density. However, azimuthal variations, like spiral arms, rings, are also seen. We use the shapelet decomposition of the specific intensity  to model its large-scale structure. Shapelets are defined as a set of localized  basis functions with different shapes \citep{2003MNRAS.338...35R}, we use Gaussian weighted Hermite polynomials in polar coordinates here. In terms of these shapelet basis $S_{nm} (\vec{\theta}, \beta)$ and the shapelet coefficients $f_{nm}$, the specific intensity can be written as
\begin{equation}
I(\vec{\theta}) = \sum \limits _{n=0} ^{\infty} \sum  \limits_{m=-n} ^{n} f_{nm}  S_{nm}(\vec{\theta}, \beta).
\end{equation}
Here  $\beta$ is called the shapelet scale. Different orders $n$ of shapelet coefficients represents different scales of the specific intensity with higher orders representing the smaller scale variations in it. We define the window function as
\begin{equation}
W(\vec{\theta}) = W_{0} \sum \limits _{n=0} ^{N}  \sum \limits_{m=-n} ^{n} f_{nm} S_{nm} (\vec{\theta}, \beta).
\label{eq:windef}
\end{equation}
Here $W_{0}$ is chosen in such a way that $\int W(\vec{\theta}) d \vec{\theta} = 1$. This captures the large-scale variation of the specific intensity. Consideration of choosing the parameters $N$ and $\beta$ will be discussed shortly.

\subsection{Modeling  $\bar{I}$ and $\delta I(\vec{\theta})$}
The power spectrum of the specific intensity fluctuations $\delta I (\vec{\theta})$ is defined as 
\begin{equation}
\left \langle \delta \tilde{I}^{*}(\vec{U}) \delta \tilde{I}(\vec{U}') \right \rangle = \delta_{D}(\vec{U}-\vec{U}') P(U) ,
\end{equation}
where  $U = \mid \vec{U} \mid$.  The angular brackets denote ensemble averages.  Observations in our Galaxy \citep{1983AA...122..282C,1993MNRAS.262..327G} and external dwarf and spiral galaxies \citep{2001ApJ...548..749E, 2006MNRAS.372L..33B, 2009MNRAS.398..887D, 2013NewA...19...89D, 2013MNRAS.436L..49D} suggest that the \HI specific intensity fluctuations can be modelled as a  Gaussian random distribution having a power law power spectrum, i.e, $P(U) \propto U^{\alpha}$. We use the  parameters $\alpha$ and $\mathcal{R}$, the ratio  of total intensity $\bar{I}$ to the standard deviation of these fluctuations $\sigma_{\delta I}$  to simulate zero-mean Gaussian random numbers with a given power law power spectrum to represent $\delta I(\vec{\theta})$.

\begin{figure}
\begin{center}
\includegraphics[scale=.5]{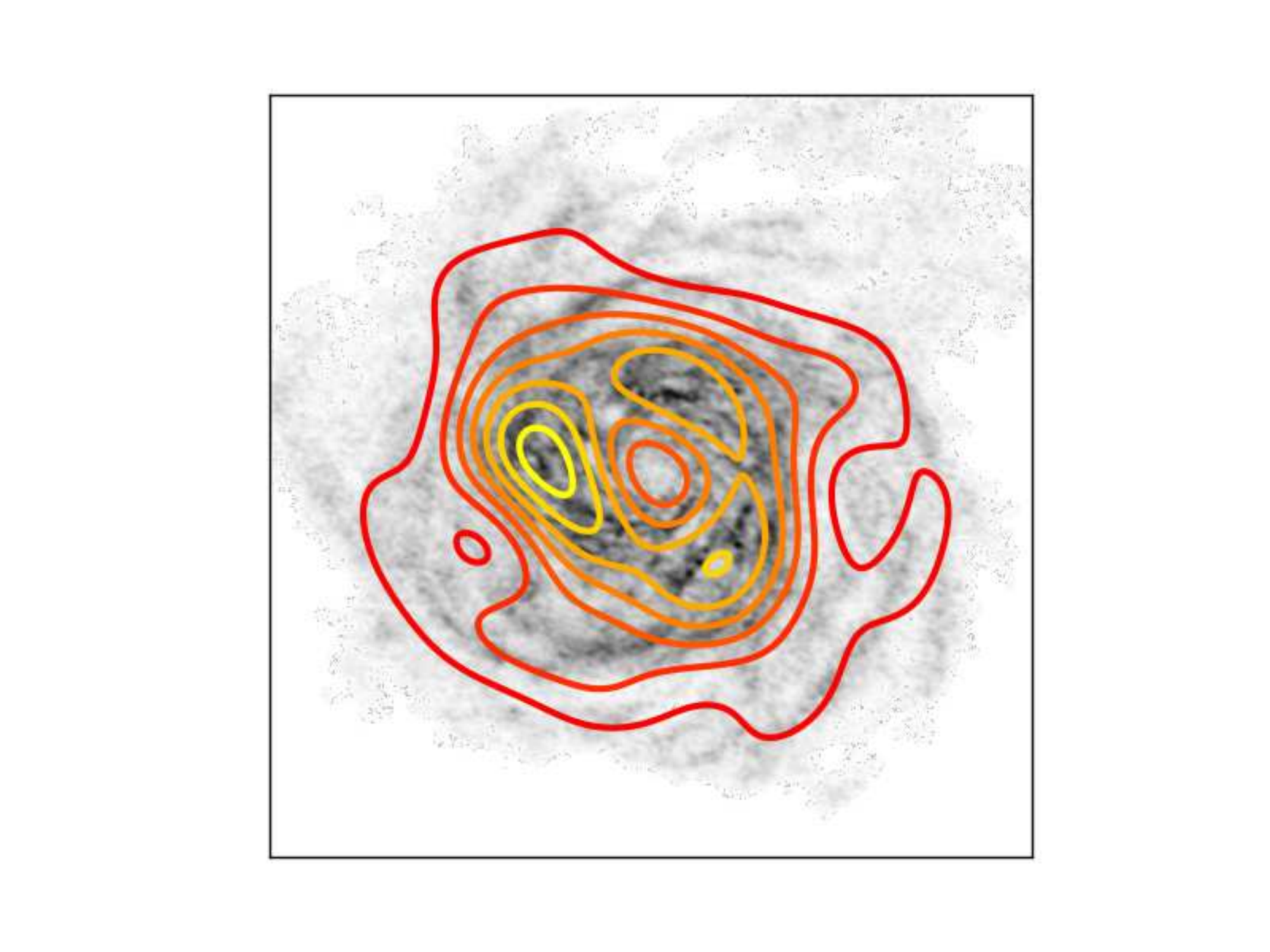}
\caption{Greyscale image showing  the natural weighted moment zero map of the galaxy NGC~628. Contours represent the model window function based on this map (see eqn~\ref{eq:windef}).}
\label{fig:win}
\end{center}
\end{figure}

\subsection{Modeling the specific intensity distribution $I(\vec{\theta})$}
We  model the window function based on the large-scale structure of the face-on spiral galaxy NGC~628. We decompose the moment zero map (natural weighted) of NGC~628 taken from THINGS \citep{2008AJ....136.2563W} survey data product \footnote{THINGS: The HI Nearby Galaxy Survey\\ data product: \url{http://www.mpia.de/THINGS/Data.html}} in terms of its shapelet coefficients and use the first few shapelets to model the window function. We choose the largest shapelet order $N$ and the shapelet scale $\beta$ as follows. Considering a given value of $\beta$, we construct the zeroth order shapelet ($N=0$, Gaussian function ) from the moment zero map of NGC~628 and estimate the mean square difference between the moment zero map and this basic shapelet. The lowest mean square difference corresponds to $\beta = 240''$ for the galaxy NGC~628.  \citet{2013NewA...19...89D}   found  that the intensity fluctuations in the  galaxy NGC~628 are dominated by the window function at angular scales $< 240''$.   We found that for $N<=12$ the shapelet coefficients do not have significant structures at angular scales $<240''$. Hence we use $\beta=240''$ and  $N = 12$ to  construct the model window function. The Grey scale image in the figure~(\ref{fig:win}) represents the moment zero map (natural weighted) of NGC~628 from THINGS archive.  We show the model window function based on this moment zero map as red contours in the same figure. 
\begin{figure}
\includegraphics[scale=.5, trim={0.5cm 0. 2cm 0.}, clip]{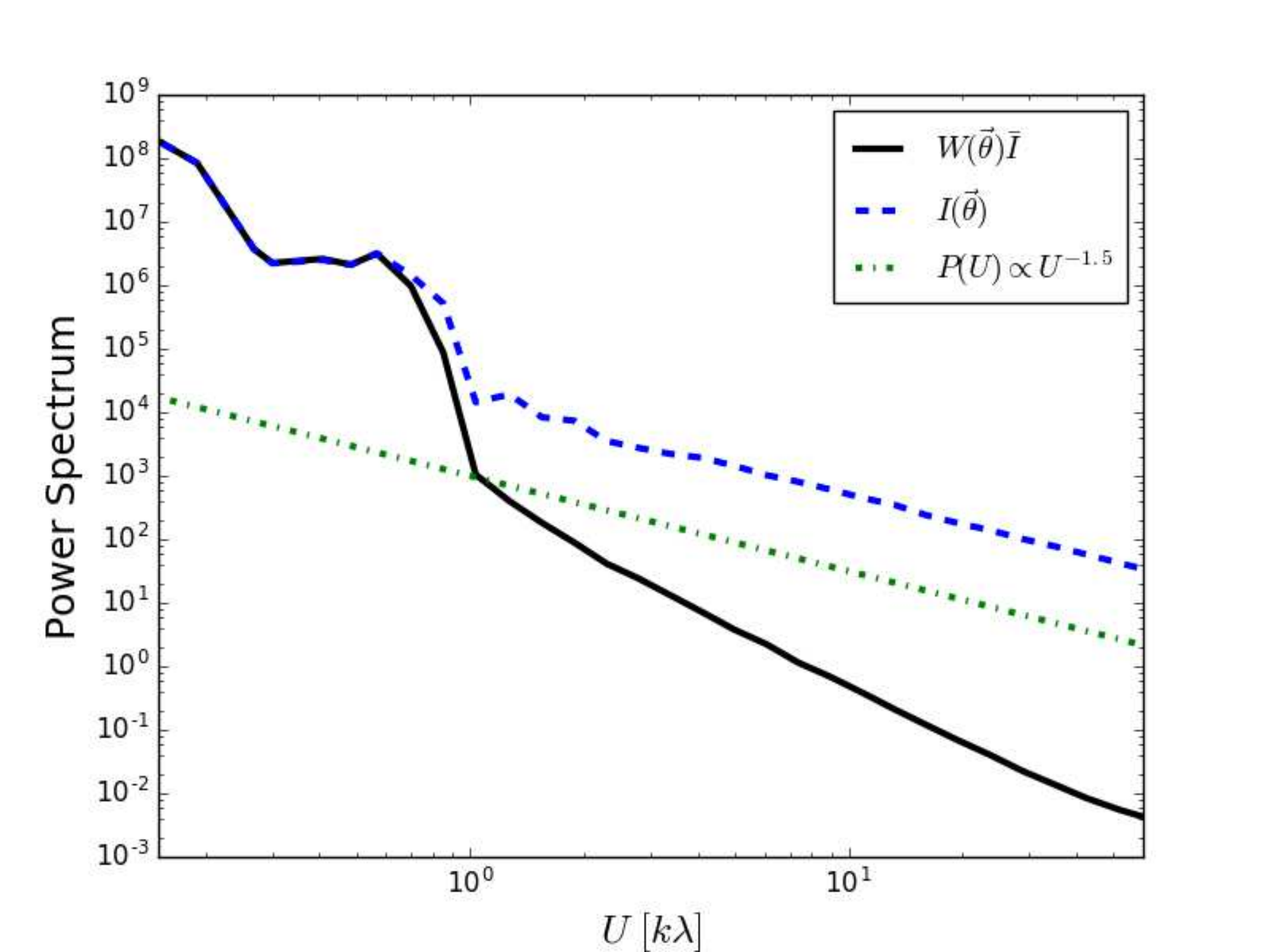}
\caption{Power spectrum for the first term $W(\vec{\theta}) \bar{I}$ in eqn.~(\ref{eq:mod}) (solid black line) is compared with the power spectrum of $I(\vec{\theta})$ for the model image (blue dashed line) with $\mathcal{R} = 5$ and $\alpha = -1.5$. The green dot-dash line corresponds to a power law with index $-1.5$.}
\label{fig:winpow}
\end{figure}
Figure~(\ref{fig:winpow}) shows the power spectrum of a model sky brightness distribution $I(\vec{\theta})$ (eqn~\ref{eq:mod}) in blue dashed line. The values of the model parameters are $\mathcal{R}=5,\ \alpha = -1.5$. The green dot-dashed line shows a power law of slope $-1.5$. Power spectra of only the first term $W(\vec{\theta}) \bar{I}$, is shown with a black solid line. Clearly, for baselines greater than $1\ k\lambda$, the power spectrum of the model image follows a power law with $\alpha = -1.5$, whereas at shorter baselines it is dominated by the effect of the window function. 

 \citet{2013NewA...19...89D} has estimated the power spectra of 18 spiral galaxies from the THINGS sample using a visibility based estimator. They found that the power spectra follow power laws with  $\alpha$  ranging  between $-0.3$ to $-2.2$. Moreover,   9 of the 18 galaxies in their sample have $\alpha$ in between $-1.5$ to $-1.8$. We choose three values of $\alpha$ for our model sky image : $[-0.5, -1.5, -2.0]$. \citet{2013MNRAS.436L..49D} found that the $\mathcal{R}$ varies between $5$ to $10$ for the six galaxies they have analyzed. We consider two values of $\mathcal{R}$ here: $[5,10]$. 

Using the above parameters we generate six model sky specific intensity distributions  in a square grid of $1024^{2}$ with each grid element representing a $1.5^{''} \times 1.5^{''}$ patch in the sky. 

\subsection{Simulated visibility data}
\begin{figure}
\begin{center}
\includegraphics[scale=.5]{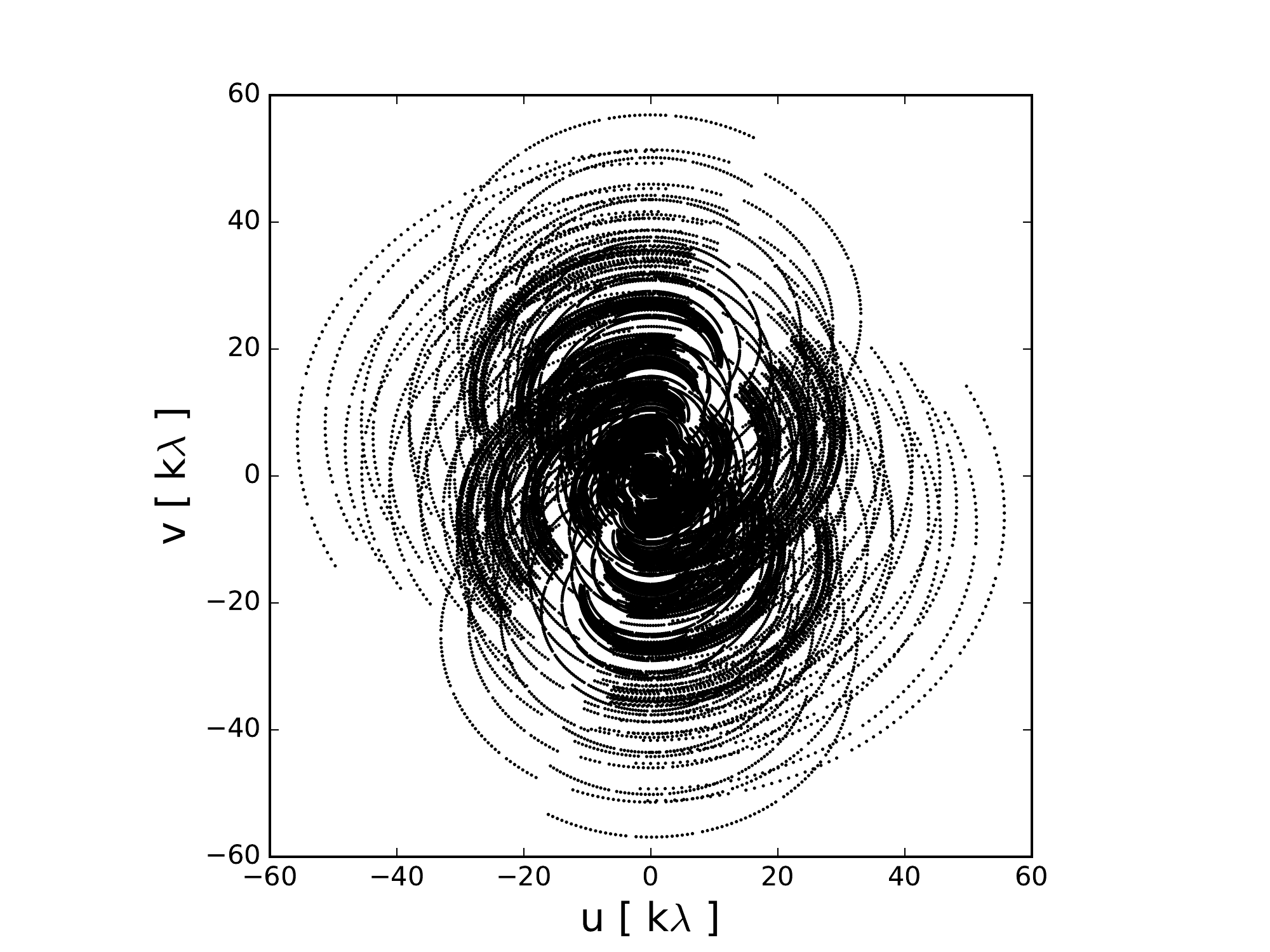}
\caption{Figure showing the sampling function for the simulated observation presented here. Black points are the places in the baseline plane where the visibilities are measured.}
\label{fig:sampling}
\end{center}
\end{figure}

\begin{figure}
\begin{center}
\includegraphics[scale=.45]{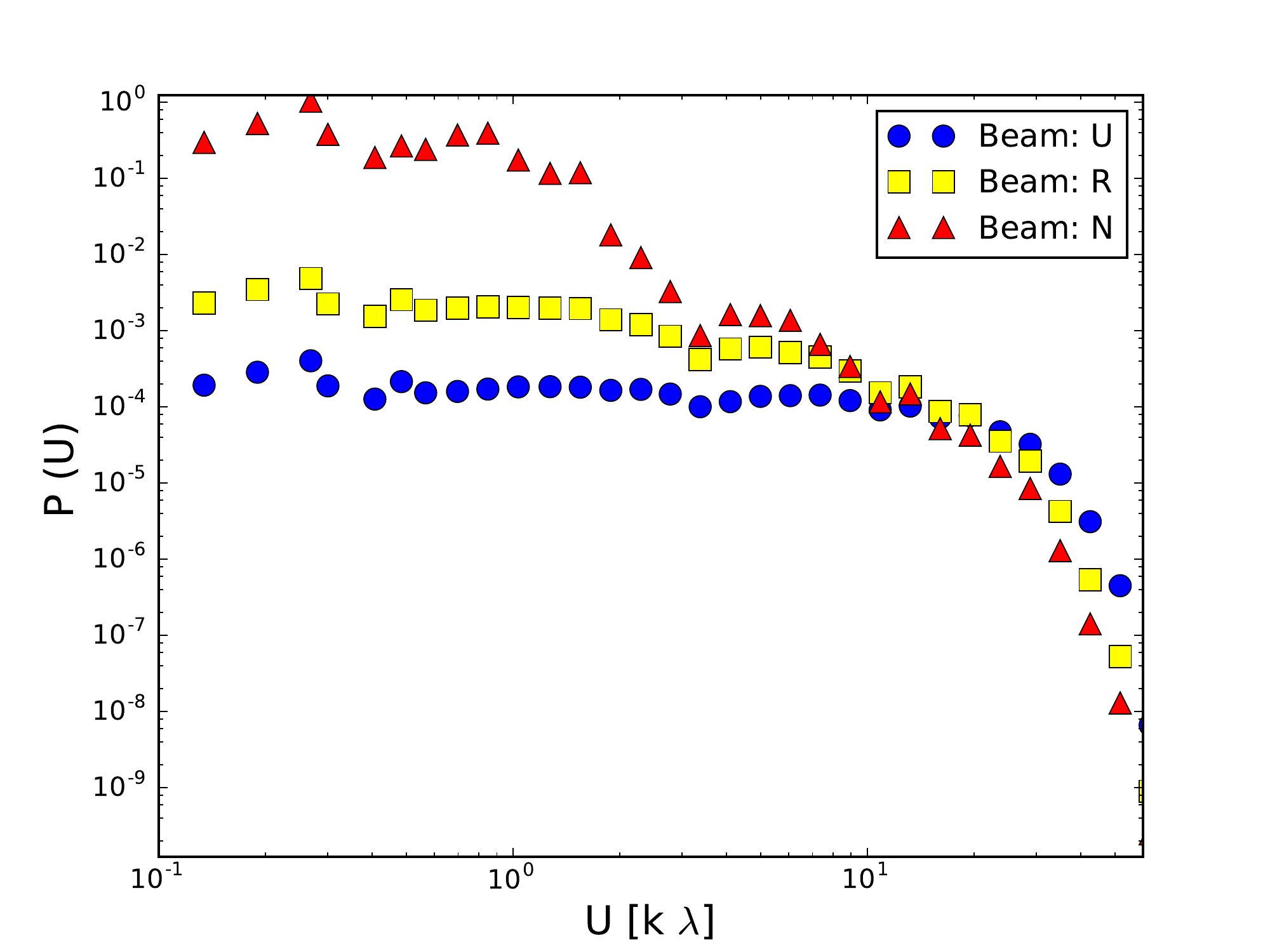}
\caption{Figure showing the power spectrum of the dirty beam for the uniform, robust and natural weighted beams for the sampling function given in Figure \ref{fig:sampling}.}
\label{fig:beamps}
\end{center}
\end{figure}

To simulate radio interferometric observations and generate random group visibilities from the above sky model, we need to choose a particular array configuration of the interferometer. We model our telescope based on the GMRT array configuration\footnote{GMRT original array configurations can be seen in \url{http://www.gmrt.ncra.tifr.res.in/gmrt_hpage/Users/doc/obs_manual}}. We scale the antenna coordinates to half its original values. This decreases the largest baseline available for the array   to $60\ k\lambda$ (instead of $\sim 120\ k\lambda$ for the original GMRT array configuration) at 21 cm and hence also reduces the effective resolution of the array. Note that, this compromise is made to  increase the computational speed and does not limit our analysis of assessing the efficacy of different estimators. We choose the declination of the source to be $+54^{\circ}$, which produces a fairly good uv-coverage. For each of the six model specific intensity distributions,  we perform eight hours equivalent of simulated observation. Figure~\ref{fig:sampling} show the sampling function corresponding to this simulated observation. We also show the power spectrum of the dirty beam corresponding to the three different weighting schemes in figure~\ref{fig:beamps}. Clearly, the power spectra of the three beams are significantly different. We use discrete Fourier transform to calculate the complex visibility values at the sampled baseline positions in the uv plane from the model sky images.  In principle,  the measurement noise at each baseline as well as its effect  in the one and two point statistics can be arbitrarily  decreased by increasing the total observation time.  Hence, we do not include any measurement noise in our model observations for simplicity.  

\section{Different statistical estimators}
\subsection{One Point statistics: large-scale distribution of specific intensity}
The large-scale distribution of \HI  carries important information about its interplay with  star formation  in spiral and dwarf galaxies. Most  spiral galaxies show visible depression of \HI near the central part owing to large star formation rate \citep{2014MNRAS.441.2159W}. \citet{1538-3881-141-1-23} have observed the   presence of \HI holes at small knots of star formation in the disks of the spiral galaxies. Dwarf galaxies like GR~8 have a clear spatial correlation between star formation rate and \HI column density \citep{2003AA...409..879B}.  These studies require evaluating the  locally averaged   \HI intensity distribution from the observed visibilities.  An estimate of the window function can be achieved by performing local average of the  specific intensity as $\langle I(\vec{\theta}) \rangle = \bar{I} W(\vec{\theta})$.

We reconstruct the specific intensity distribution corresponding to the observed visibilities using CLEAN as discussed in the previous section. In this work, we explore various user-defined parameters with CLEAN to assess their effect on the reconstructed image. These will be discussed in detail in the next section. For each reconstructed image, we evaluate the window function using  its shapelet coefficients. Following the same arguments as  discussed in section~3.3, we use the shapelet scale to be $240''$  and   the first 12 shapelet coefficients to represent the window function. To distinguish the window function estimated from the reconstructed image from the model window function, we mention the earlier by $W_{C}(\vec{\theta})$ for the rest of the analysis. Further, for the model ($W(\vec{\theta})$) as well as the reconstructed windows ($W_{C}(\vec{\theta})$),  we estimate the azimuthally averaged window function defined as 
\begin{equation}
W_{A} (\theta) = \frac{1}{2 \pi} \int \limits _{0}^{2 \pi} W(\theta, \phi) d \phi,
\end{equation}
in different bins of $\theta$. Here $\left ( \theta, \phi \right )$ are the polar components of the vector $\vec{\theta}$. We use the standard deviation of the values of the reconstructed window  $W_{C}(\vec{\theta})$ in each azimuthal bin to represent the  statistical fluctuations associated with the estimated value of $W_{A} (\theta)$ in the corresponding bin. This gives us a robust way of comparing the estimates of the window function with that of the model.

\subsection{Two point statistics: Power spectrum}
Two-point statistics of any field quantifies the scale dependence of fluctuations in it. There are several quantifiers of the two point statistics. For a two dimensional  field  $A(\vec{\theta})$, the structure function and the autocorrelation  function evaluates the two-point statistics in the $\vec{\theta}$ plane while the power spectrum of it evaluates the two-point statistics in a plane Fourier conjugate to $\vec{\theta}$, like the baseline plane $\vec{U}$ (see \citet{2004ARAA..42..211E} for more detailed analysis of these estimators). For a Gaussian random field, all these different estimators contain the same information. We restrict  ourselves in measuring the power spectrum of the sky brightness fluctuations, i.e the quantity $\delta I(\vec{\theta})$ here. As discussed before, power spectrum estimators from the interferometric data can be categorized into two classes the {\em image based estimator} and the {\em visibility based estimator}. We give a brief description of these estimators here.
\subsubsection{Image based power spectrum estimator}
Image based estimators use the reconstructed image to estimate the power spectrum.  To distinguish the reconstructed image from the sky brightness distribution, we shall denote the former by $I_{C}(\vec{\theta})$. Since the image is already evaluated at regular grids in $\vec{\theta}$, a two dimensional FFT can be used to estimate the Fourier transform of 
$I_{C}(\vec{\theta})$. As the interferometers are mostly not sensitive at baselines lower than a certain value, they effectively do not measure the first term in the eqn~(\ref{eq:mod}) and we may write 
\begin{equation}
\tilde{I}_{C}(\vec{U}) = \tilde{W}(\vec{U}) \otimes \delta \tilde{I}(\vec{U}) + \mathcal{B}_{I}(\vec{U}),
\end{equation}
where $ \tilde{W}(\vec{U})$ represents the Fourier transform of the window function and  $\otimes$ denotes the convolution. The quantity $\mathcal{B}_{I}(\vec{U})$ jointly represents any artifacts introduced in the image reconstruction procedure and effective noise in $\tilde{I}_{C}(\vec{\theta})$ resulting from the measurement noise. We correlate $\tilde{I}_{C}(\vec{U})$ at each baseline, which gives 
\begin{equation}
P_{C}(\vec{U}) = \langle \tilde{I}_{C}(\vec{U}) \tilde{I}^{*}_{C}(\vec{U}) \rangle = \mid  \tilde{W}(\vec{U})  \mid^{2} \otimes P(\vec{U}) + P_{\mathcal{B}} (\vec{U}),
\label{eq:imps}
\end{equation}
 where $P_{\mathcal{B}} (\vec{U})$ is related to both $\mathcal{B}_{I}(\vec{U})$ and $\tilde{I}(\vec{U})$. The angular brackets above denotes the ensemble average of many realizations of the sky. In practice, we assume statistical isotropy and choose  azimuthal bins  to perform this average.  Hence the image based azimuthally averaged power spectrum estimator is given as 
\begin{equation}
P_{I}(U) = \frac{1}{2 \pi} \int \limits_{0} ^{2 \pi} \tilde{I}_{C}(U, \phi) \tilde{I}^{*}_{C}(U, \phi )  d \phi,
\end{equation}
where $(U, \phi)$ are the polar components of the vector $\vec{U}$. As discussed before (see figure~\ref{fig:winpow}), the window function represents large-scale variation of the specific intensity and hence at baselines $U >> 1/\theta_{0}$, where $\theta_{0}$ is the extent of the galaxy, the window function can be treated as a delta function $\delta_{D} (\vec{U})$. Hence in the absence of  $P_{\mathcal{B}}$, at baselines $U >> 1/\theta_{0}$, the quantity $P_{C}(\vec{U})  \sim P(\vec{U})$ and $P_{I}(U)$ gives an estimate of the power spectrum of the sky brightness fluctuations. The quantity $P_{\mathcal{B}} (\vec{U})$ is a manifestation of the incomplete baseline coverage and the different techniques incorporated in the image reconstruction process. Arguably it depends on the user chosen parameters in the CLEAN (or other algorithms) and hence need to be evaluated and subtracted from the above equation to estimate the power spectrum in an unbiased way. Unfortunately, a separate estimation of $P_{\mathcal{B}} $ is almost always impossible. Interestingly, such a bias is grossly ignored in literature where image based estimators are used \citep{2012ApJ...754...29Z, 2014MNRAS.441..525W}. Further,  incomplete baseline coverage also makes the measurement  noise correlated in $\tilde{I}_{C}(\vec{\theta})$ and introduces a non zero $P_{\mathcal{B}} (\vec{U})$ (see \citet{2011arXiv1102.4419D} for detail). However, as we have not considered the measurement noise in our simulation, we refrain from investigating this effect here.   Estimating the errors in the reconstructed image is not straightforward and only a Monte-Carlo based technique can be effectively used \citep{1999ASPC..180..419S}. Following that, the errors in the image based estimates of the power spectrum are also non-trivial. We use the variation of the power spectrum values in different $\vec{U}$ inside an annular bin to represent the error in the image based power spectrum estimator. Additionally, at smaller baselines  independent estimates of the power spectrum reduce and the sample variance dominates. The sample variance is given by $P_{I}/\sqrt{N_{g}}$, where $N_{g}$ is the number of independent estimates of $P_{I}$ in a given annular bin. We also add this in quadrature to represent the error in the image based estimator of the power spectrum.

\subsubsection{Visibility based power spectrum estimator}
The visibility based power spectrum estimators use the directly measured visibilities and do not require  image reconstruction \citep{2001JApA...22..293B}.  Since the Fourier transform of the first term in eqn.~(\ref{eq:mod}) is not measured mostly by an interferometer (see the discussion above), the measured visibilities can be written as 
\begin{equation}
V(\vec{U}) = \left[  \tilde{W(\vec{U})} \otimes \delta \tilde{I}(\vec{U}) \right] S(\vec{U}) + \mathcal{N} (\vec{U}).
\end{equation}
The visibility correlation gives
\begin{equation}
\langle V(\vec{U}) V^{*}(\vec{U}) \rangle = \mid  \tilde{W(\vec{U})} \mid^{2} \otimes  P(U) \mid S(\vec{U}) \mid^{2} + \mid \mathcal{N} (\vec{U}) \mid^{2}.
\end{equation}
In absence of measurement noise, at baselines $U >> 1/\theta_{0}$, the visibility correlation gives $P(U) \mid S(\vec{U}) \mid^{2} $. The azimuthally averaged power spectrum then can be estimated as
\begin{equation}
P_{V}(U) = \int \limits_{0} ^{2 \pi} \langle V(\vec{U}) V^{*}(\vec{U}) \rangle  d \phi / \int \limits_{0} ^{2 \pi} \mid S(\vec{U})\mid^{2}   d \phi.
\end{equation}
In practice, it is estimated at discrete azimuthal bins. For most of the array configurations, the integral in the denominator of the above expression has a nonzero value. However, if the integral is zero in a particular bin the power spectrum is not evaluated at that bin. In realistic observations, the noise term $\mid \mathcal{N} (\vec{U}) \mid^{2}$ dominates and introduces a bias in power spectrum estimates. \citet{2011arXiv1102.4419D} discusses the procedure to take care of this noise bias in detail. Since we do not have measurement noise in our simulation, we neglect this effect here. We estimate the errors in the visibility based power spectrum estimator following the calculations by \citet{2011arXiv1102.4419D}.

\section{Analysis  and Result}

\begin{figure*}
\includegraphics[scale=.41, clip]{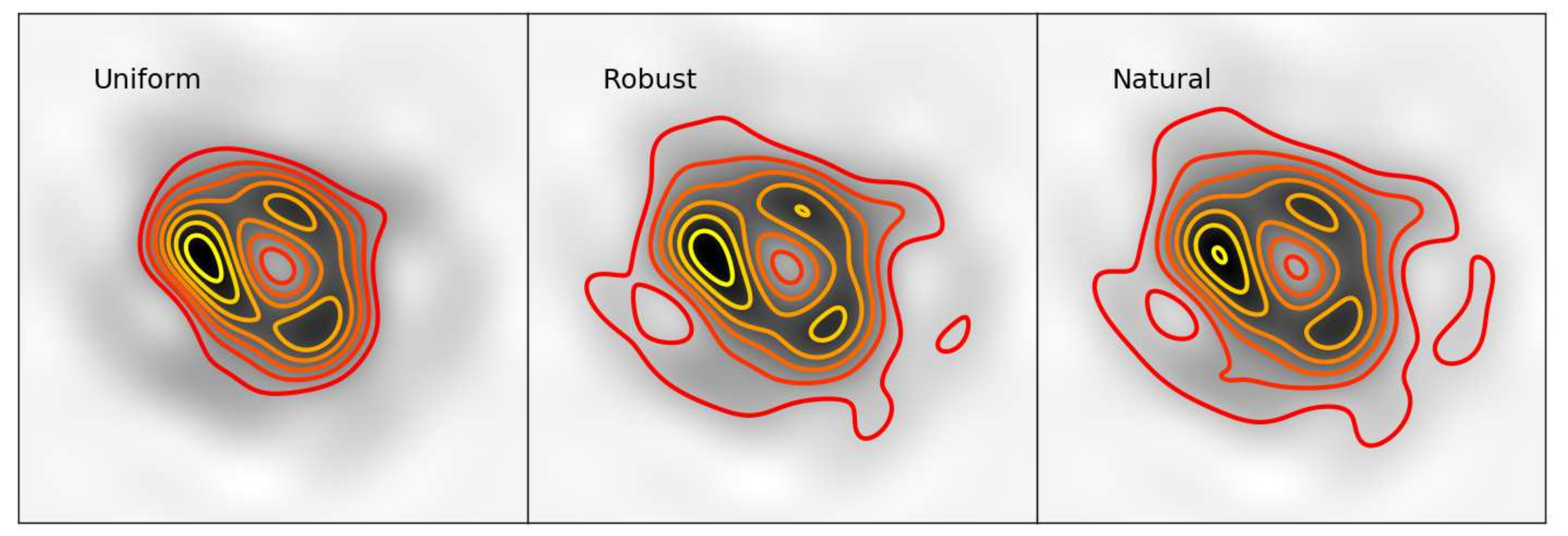}
\caption{Window functions of the reconstructed images using different weighting schemes are shown in contours against the grey scale plot of the window function of the model image corresponding to  $\mathcal{R} = 5,\ \alpha = -1.5$.}
\label{fig:win3w}
\end{figure*}

\begin{figure}
\includegraphics[scale=.46, clip]{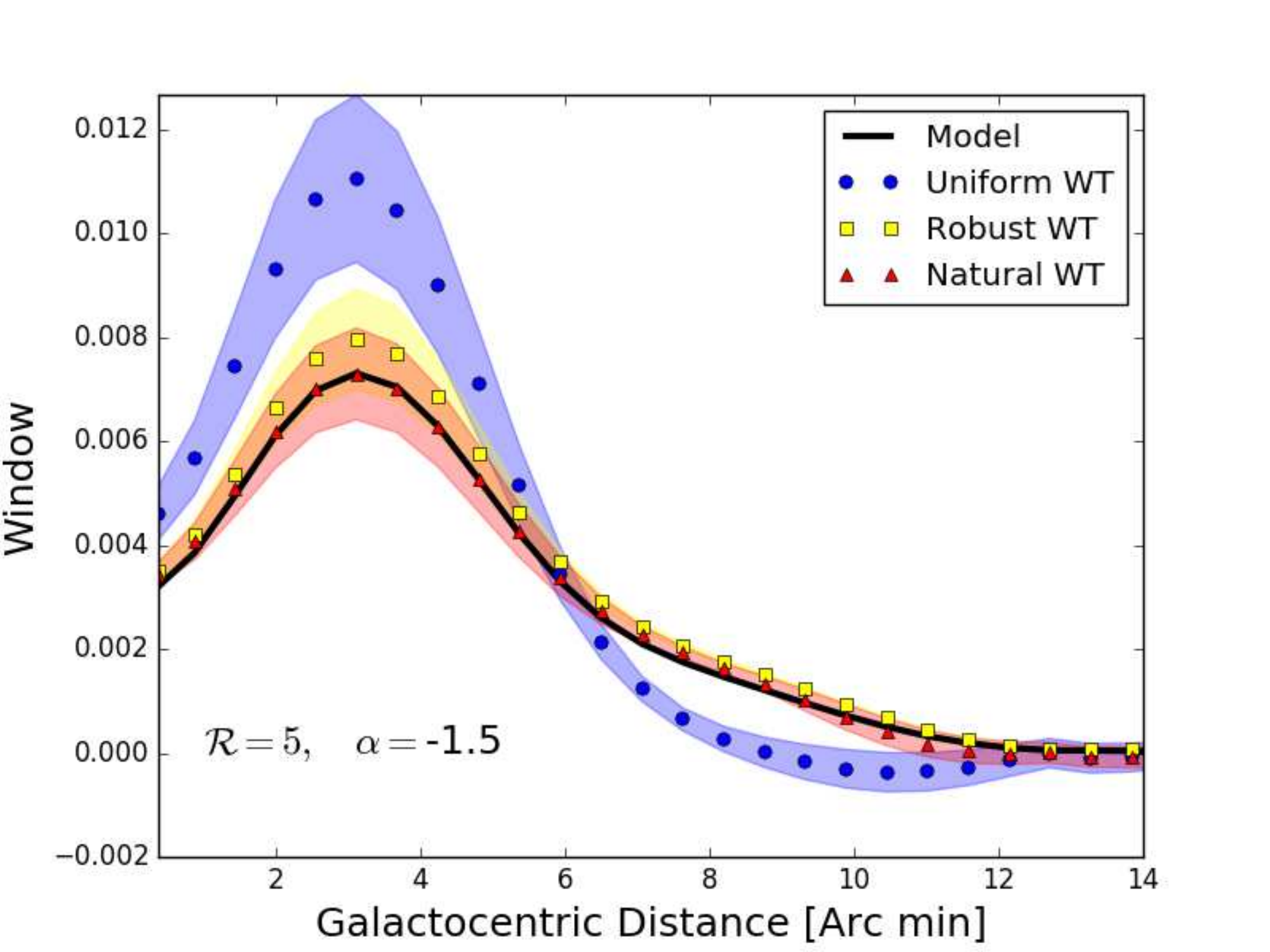}
\caption{This figure compares the azimuthally averaged window functions estimated using different weighting schemes with that of the input image. The colour bands for each estimate represents the error.}
\label{fig:win8}
\end{figure}

We use the task IMAGR in  AIPS   to reconstruct the sky brightness distribution for each of the six simulation sets. We discuss  our analysis and results based on the simulation with parameters $\mathcal{R} = 5.0, \alpha=-1.5$ in detail and tabulate the results for all the models. 

In choosing different parameters for CLEAN in the task IMAGR, we give particular emphasis on the fact that here we are interested in reconstructing the sky brightness distribution for  diffuse emission. It is a common understanding that a smaller loop gain improves the reconstruction of extended sources \citep{2017isra.book.....T}, however improvement for a gain $<0.01$ is minimal. We choose a loop gain of $0.005$. 
To tame the effect of abrupt fall in the baseline coverage at large baseline we use a Gaussian taper with the UVTAPER parameter set to 45 k$\lambda$ in IMAGR ( for both $u$ and $v$). This corresponds to a tapering function  $T(U)  = exp\left [ - U^{2} / 2 (30)^{2} \right] $, where $U$ is measured in k$\lambda$.  Tapering down-weights the visibilities at the larger baselines and hence may have an effect on the power spectrum estimates.  However, for a power law power spectrum the effect of tapering can be analytically reversed by multiplying the power spectrum by $1/T(U)^{2}$. We choose the pixel size for the image to be $1.5^{''} \times 1.5^{''}$ in a grid of $1024^{2}$.  We use three different weighting schemes to weight and grid the visibility data, namely the natural weighting, the uniform weighting and the robust weighting. These are controlled mainly by the parameters ROBUST in AIPS. We have chosen ROBUST values of $(-5, 0, 5)$ to produce three different reconstructions of sky brightness from each simulated visibility dataset. For each IMAGR run, we manually stop the major cycles when the maximum and minimum pixels in the residual image are of similar value. The restoring beams for the uniform, robust and natural weighted images came out to be $4.1'' \times 3.9''$, $6.7'' \times 5.8''$ and $9.3''\times 8.2''$ respectively with the beam position angles $\sim 63^{\circ}, \sim 68^{\circ}$ and $\sim 70^{\circ}$. These images are  used for further analysis.

\subsection{Window function}
For each of the reconstructed image, we estimate the window function $W_{C}$ and the azimuthally averaged profile $W_{A}$ following the prescriptions given in the previous section. Contours in three panels of figure~(\ref{fig:win3w}) show $W_{C}$ corresponds to the uniform, robust and natural weighting schemes respectively. The Greyscale image in each panel corresponds to the model window function. A visual comparison of the contours of these three panels with those in  figure~(\ref{fig:win}) demonstrates that the natural weighted scheme best reproduces the model window. We shall quantify this statement shortly. Note that both the model window function $W$ and the quantity $W_C$ are estimated at Cartesian grids in $\vec{\theta}$, we may denote the values in the grid points as $W[i,j]$ and $W_{C}[i,j]$ respectively. We use the quantity
\begin{equation}
\chi = \frac{\sum \limits_{i, j} \left ( W[i, j] - W_{C}[i, j] \right ) ^{2}}{N_{G}^{2}\sqrt{ \sigma_{W}^{2} + \sigma_{C}^{2}}}
\label{eq:defchi}
\end{equation}
to represent the deviation of  $W_{C}$ from $W$. Here $\sigma_{W}$  and $\sigma_{C}$ correspond to the standard deviation of the pixel values of $W[i,j]$ and $W_{C}[i,j]$ respectively and $N_{G}$ gives the number of grid points along one axis \footnote{The quantity $\chi$ used here has no probabilistic interpretation and should not be confused as a functional used in the most maximum likelihood estimations.} The quantity $\chi$ gives the mean square deviation between the model and the estimated window functions.  Clearly, a lower value of $\chi$ corresponds to a better reconstruction of the window function. We have performed tests to check if the figure of merit $\chi$ actually captures the deviation between the estimates of the window function. In this test, we choose a fiducial model for the window function, values of $\mathcal{R}$ and $\alpha$  and generate the different realizations of $\delta I$ to simulate models of the sky brightness distribution. We then estimate the window function from each of these images and compared them with the fiducial window function using the figure of merit $\chi$.  We found  the $\chi$ value lies in between $0.004$ to $0.01$ within the six models discussed here.   We found that the uniform weighting produces a window function with the largest value of $\chi = 2.63$, whereas the corresponding values of $\chi$ for robust weighting and natural weighting are $0.26$ and $0.15$ respectively.  The window function estimated from the images produced with uniform and robust weighting have the bias, however,  natural weighting scheme produces the best estimate of the window function.

\begin{figure*}
\includegraphics[scale=.425, trim={0.cm 0. 1.7cm 0.}, clip]{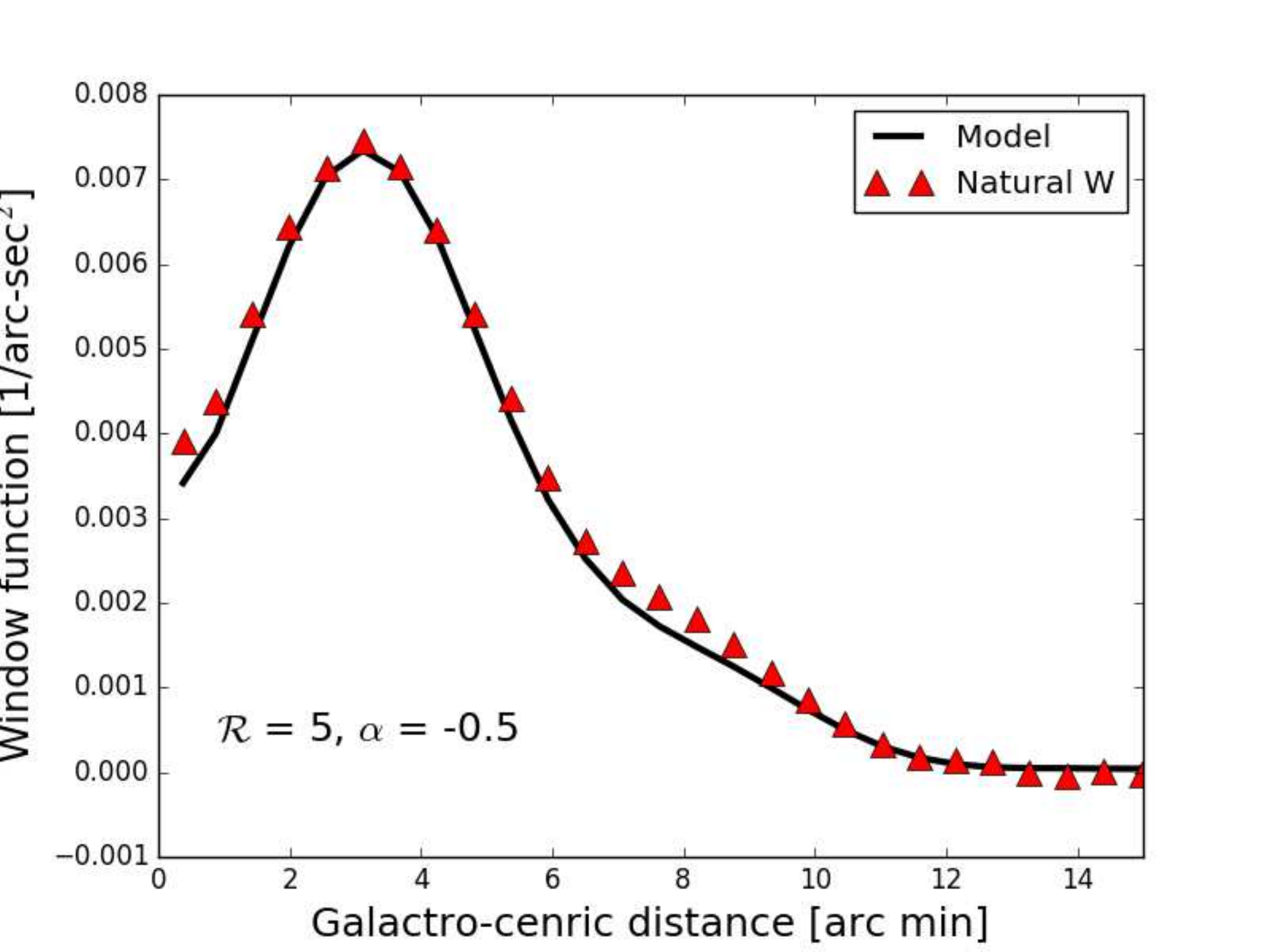}
\includegraphics[scale=.425, trim={0.6cm 0. 1.7cm 0.}, clip]{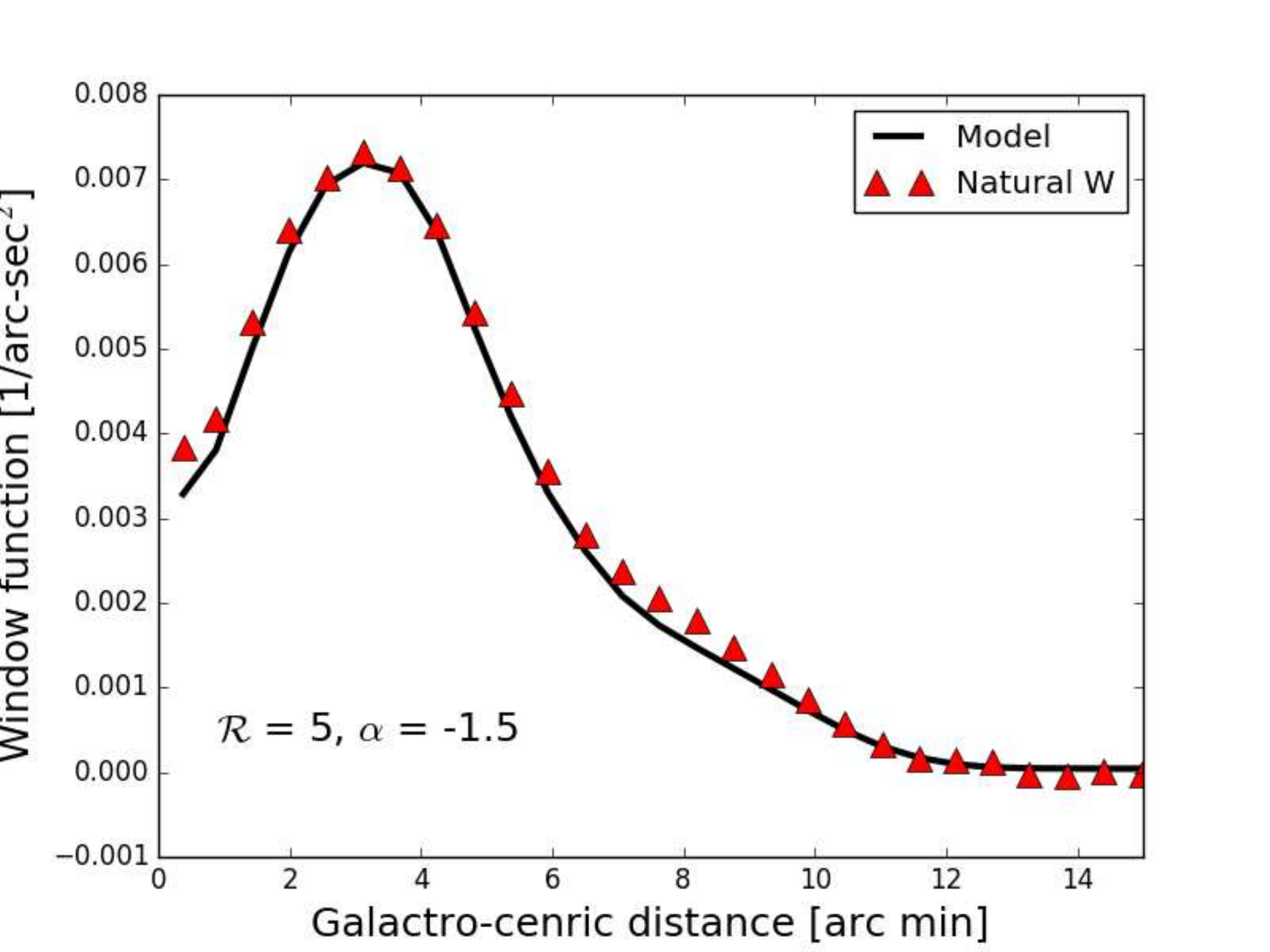}
\includegraphics[scale=.425, trim={0.cm 0. 2cm 0.}, clip]{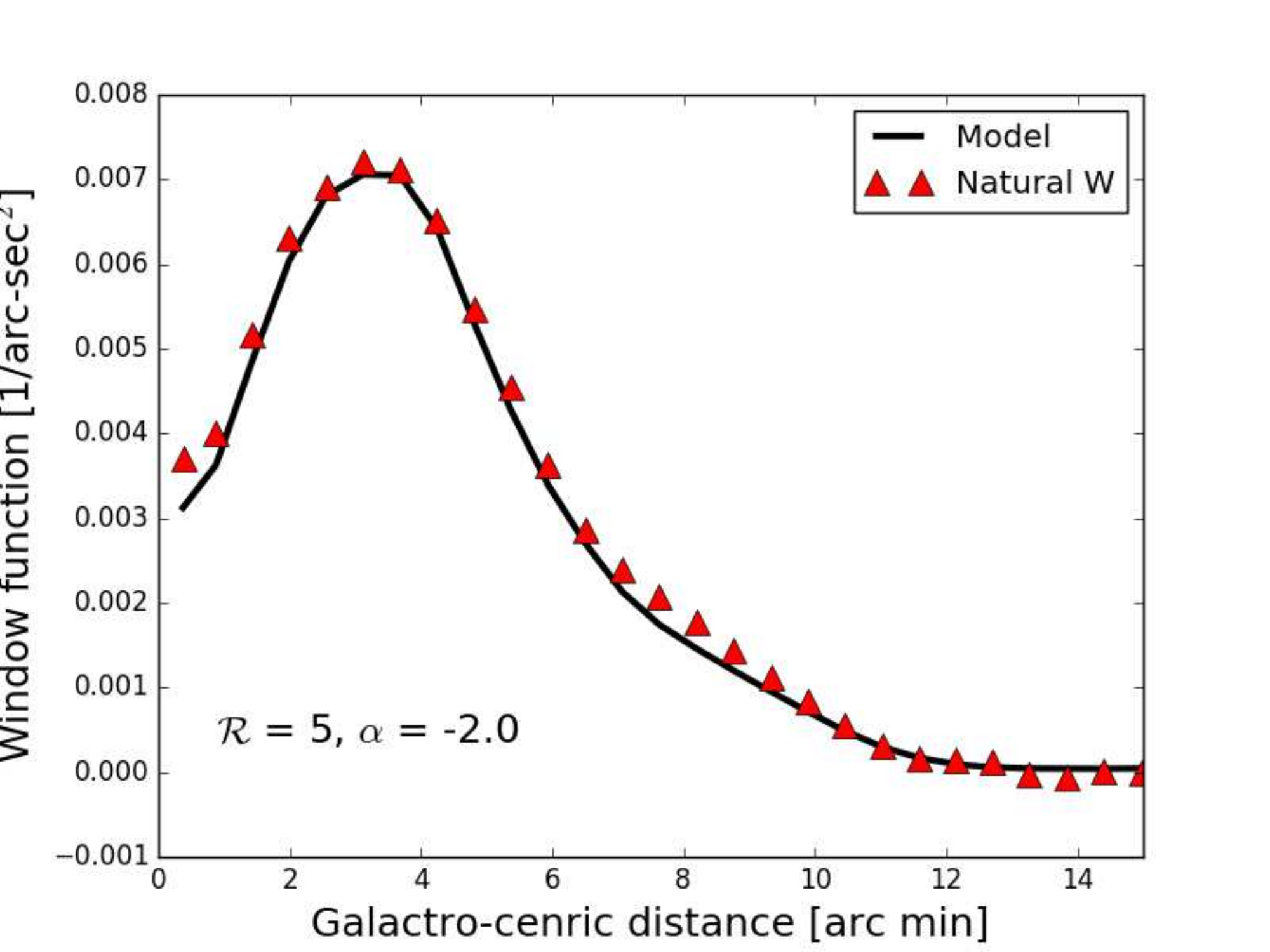}
\includegraphics[scale=.425, trim={0.6cm 0. 1.7cm 0.}, clip]{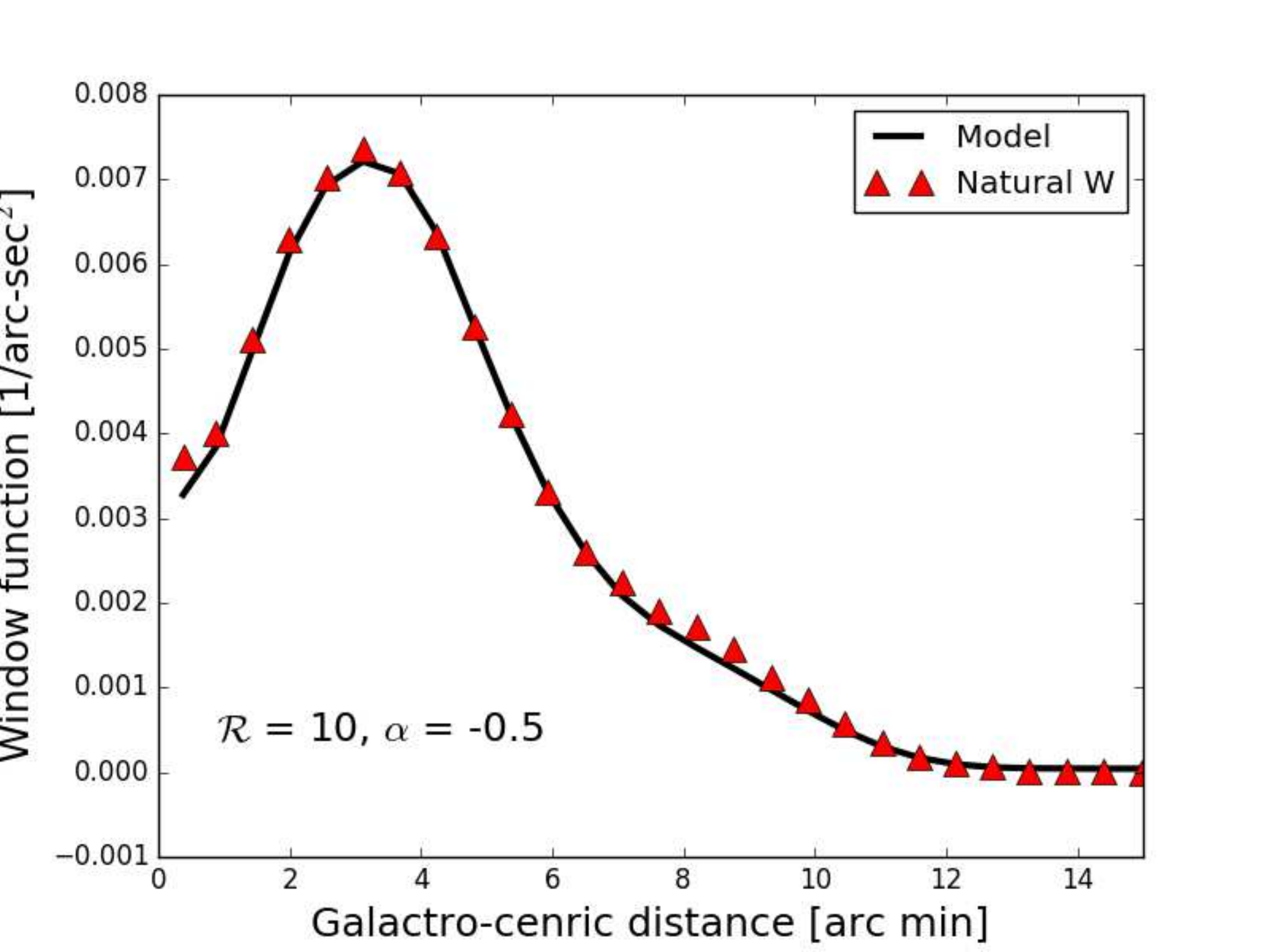}
\includegraphics[scale=.425, trim={0.cm 0. 1.7cm 0.}, clip]{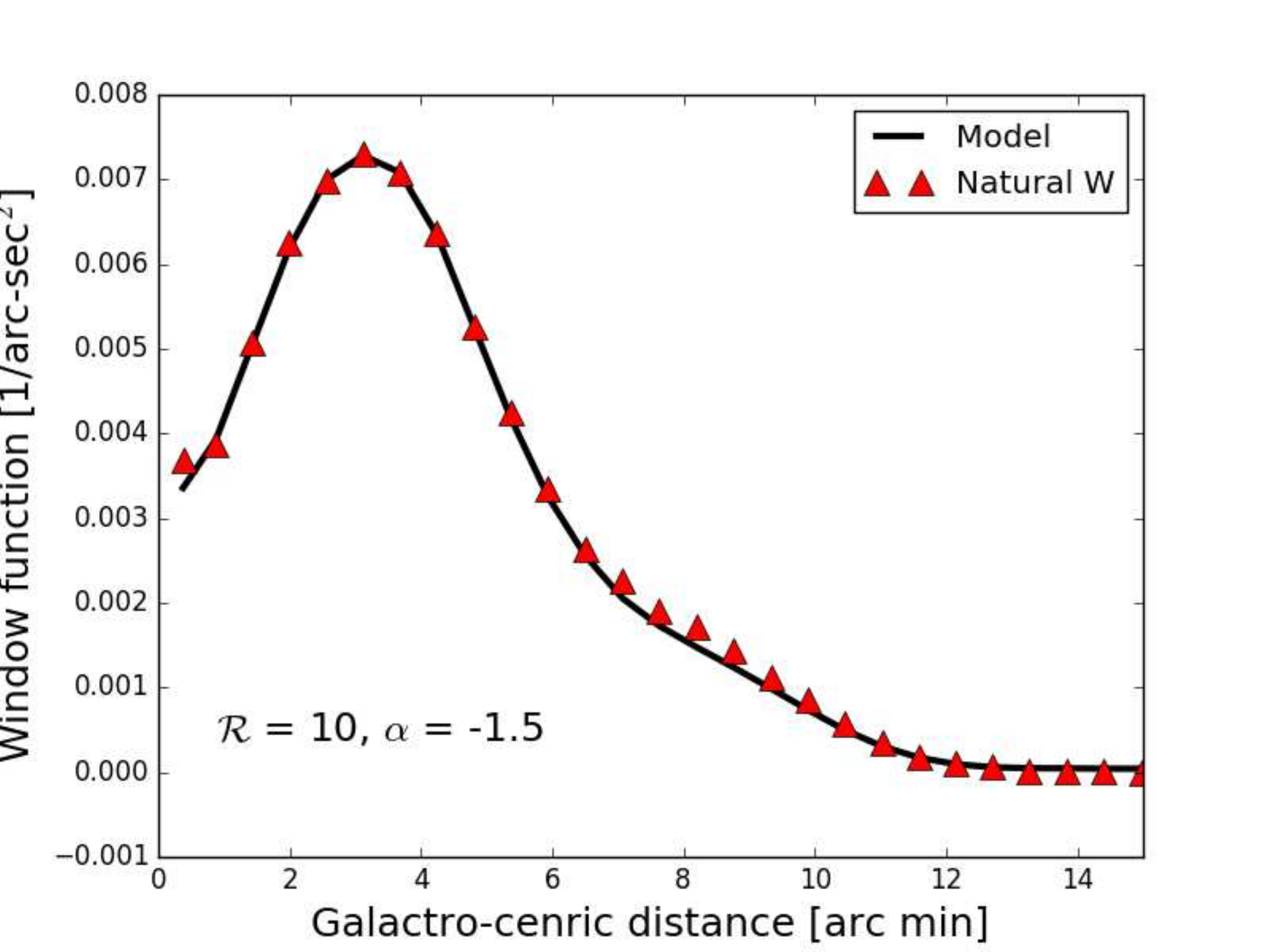}
\includegraphics[scale=.425, trim={0.6cm 0. 2cm 0.}, clip]{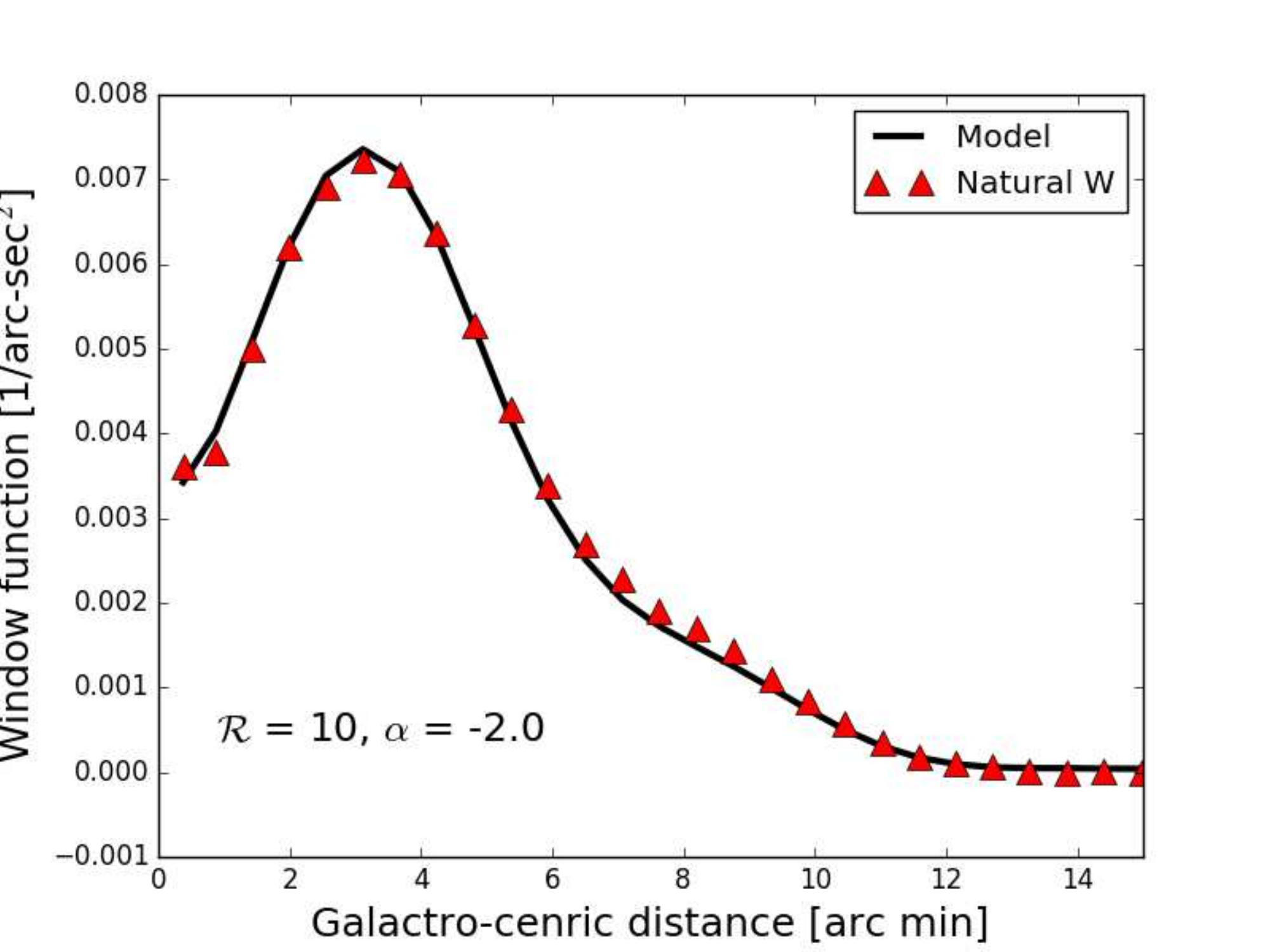}
\caption{Comparison of the azimuthally averaged window functions estimated from the input model (black solid line) and the reconstructed natural weighted image  (red triangles) are shown for all six simulations. }
\label{fig:WIN6}
\end{figure*}

Figure~(\ref{fig:win8}) plots the azimuthally averaged window functions estimated from the model as well as the three reconstructed images. The black solid line corresponds to the azimuthally averaged window function of the model image. Colour bands around each of the estimates correspond to the variation of the window function within the respective azimuthal bins. The uniform weighting (circles) produces the largest deviation from the model window and also have the largest variation in each azimuthal bins as represented by the error band in the figure. Note that both the robust (square) and the natural weighting (triangle) reproduces the azimuthally averaged window  within the errors, however, the points corresponding to the robust weighting (square) are systematically offset from the model window function. 

\begin{table*}
\centering
\begin{tabular}{|l|c|c|c|c|c|c|c|c|c|c|}
\hline
\multicolumn{2}{|l|}{}                                                                       & \multicolumn{3}{c|}{$\alpha = -0.5$}        & \multicolumn{3}{c|}{$\alpha = -1.5$}       & \multicolumn{3}{c|}{$\alpha = -2.0$}       \\ \hline
\multicolumn{2}{|l|}{}                                                                       & U            & R            & N            & U            & R            & N            & U            & R            & N            \\ \hline
\multicolumn{1}{|c|}{\multirow{3}{*}{$\mathcal{R} = 5$}} & $\chi$                            & 2.77         & 0.42         & 0.02         & 2.63         & 0.26         & 0.15         & 3.76         & 0.22         & 0.02         \\ \cline{2-11} 
\multicolumn{1}{|c|}{}                                   & $\alpha_{I}$                      & $-1.4\pm0.1$ & $-1.9\pm0.3$ & $-2.1\pm0.1$ & $-1.8\pm0.1$ & $-2.0\pm0.1$ & $-2.2\pm0.1$ & $-2.1\pm0.1$ & $-2.3\pm0.1$ & $-2.3\pm0.1$ \\ \cline{2-11} 
\multicolumn{1}{|c|}{}                                   & \multicolumn{1}{l|}{$\alpha_{V}$} & \multicolumn{3}{c|}{$-0.44\pm0.07$}        & \multicolumn{3}{c|}{$-1.5\pm0.1$}          & \multicolumn{3}{c|}{$-1.9\pm0.1$}          \\ \hline
\multirow{3}{*}{$\mathcal{R} = 10$}                      & $\chi$                            & 1.36         & 0.1          & 0.004        & 1.43         & 0.16         & 0.003        & 1.16         & 0.08         & 0.004        \\ \cline{2-11} 
                                                         & $\alpha$                          & $-1.3\pm0.1$ & $-1.4\pm0.1$ & $-1.7\pm0.3$ & $-1.9\pm0.1$ & $-2.2\pm0.2$ & $-1.9\pm0.3$ & $-2.1\pm0.1$ & $-2.4\pm0.2$ & $-2.1\pm0.3$ \\ \cline{2-11} 
                                                         & \multicolumn{1}{c|}{$\alpha_{V}$} & \multicolumn{3}{c|}{$-0.48\pm0.08$}        & \multicolumn{3}{c|}{$-1.4\pm0.2$}          & \multicolumn{3}{c|}{$-1.9\pm0.2$}          \\ \hline
\end{tabular}
\caption{Table summarising the result of comparison between different estimates of the window function and power spectrum. Efficacy to reproduce the model window function from an estimated image is quantified by $\chi$ (see eqn~\ref{eq:defchi}). We tabulate the values of $\alpha_{V}$ and $\alpha_{I}$ for each model to access the merit of different estimators in the table. The header `U', `R' and `N' corresponds to the uniform, robust and natural weighting schemes respectively.}
\label{tab:fitp}
\end{table*}

Figure~(\ref{fig:WIN6}) shows the azimuthally averaged window function estimated for all the six models using natural weighting schemes (red triangles) against the same estimated from the model images (black solid line). Table~1 shows the values of $\chi$ for all the six model skies. Clearly, in all cases, the natural weighting gives the best reconstruction of the window function. Moreover, for the models with a relatively lower amplitude of the fluctuations in specific intensity ($\mathcal{R} = 10$), the $\chi$ values are systematically lower than the models with $\mathcal{R} = 5$. We conclude, with a careful choice of the imaging parameters for CLEAN it is possible to estimate the window function unbiasedly from the reconstructed image and it is best estimated when natural weighting is used. 
\subsection{Power Spectrum}
\begin{figure}
\includegraphics[scale=.46, trim={0.5cm 0. 2cm 0.}, clip]{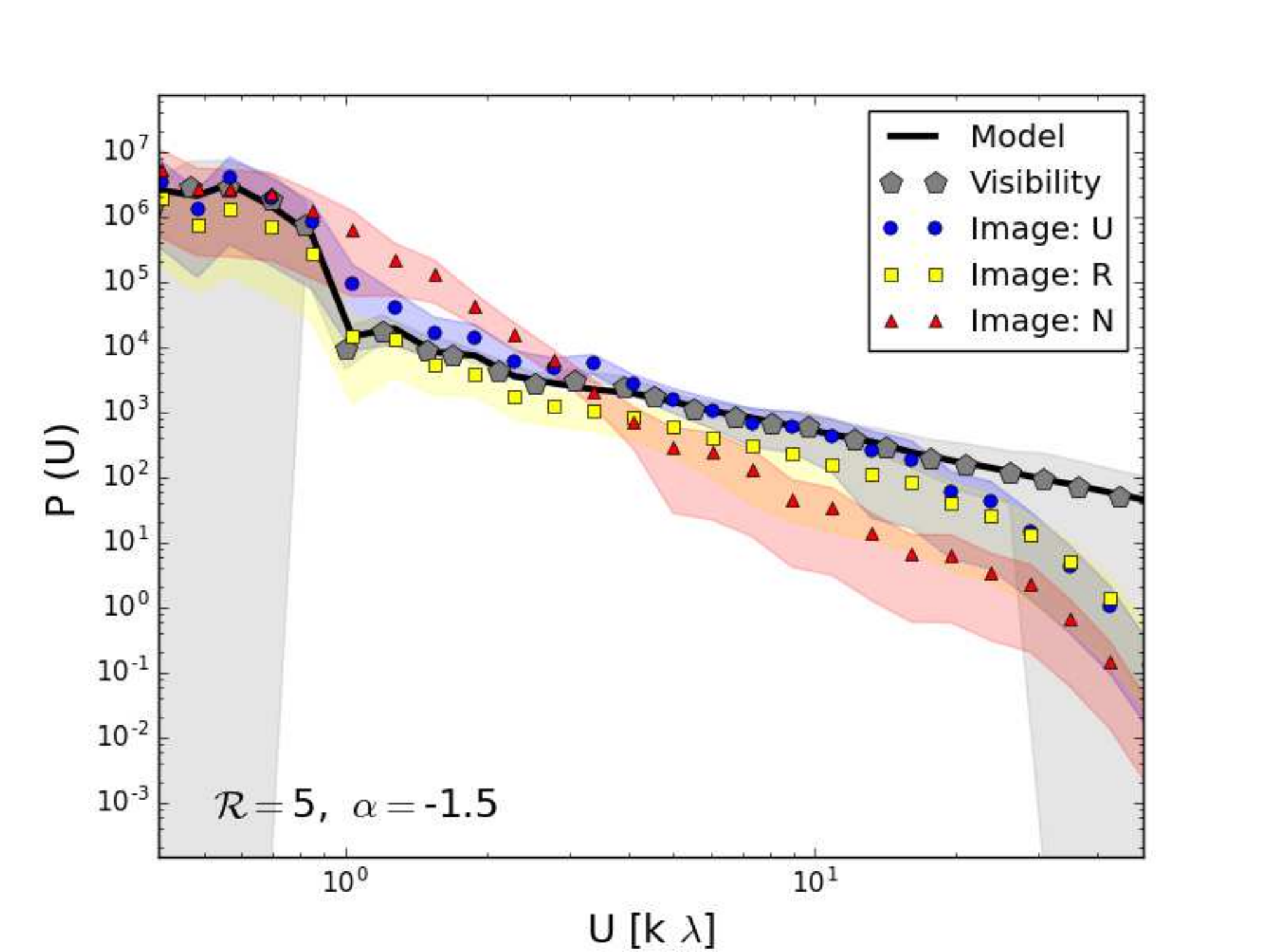}
\caption{This figure compares the power spectrum estimated using the visibility based estimator and the image based estimator for all three different weighting schemes against the model power spectrum. The shaded area corresponds to the error associated with each estimate.}
\label{fig:posp8}
\end{figure}

We estimate power spectra from the model images using the image based estimator. Note that, the model images do not have any artefacts  that may arise from the reconstruction and hence this power spectrum can be considered as a reference. This is shown with a solid black line in figure~(\ref{fig:posp8}). The power spectrum  $P_{V}(U)$ is shown with grey pentagons in the same figure with the grey area indicating the error bars. It is quite clear that the visibility based power spectrum follows the reference spectra quite well and overall bias is minimized.  The large error bars in small and larger baseline are indicative of less independent measurements at those baselines. The visibility based power spectrum estimator assumes  power law at baselines larger than $1\ k\lambda$. 
\begin{figure*}
\includegraphics[scale=.4, trim={0.2cm 0. 1.7cm 0.}, clip]{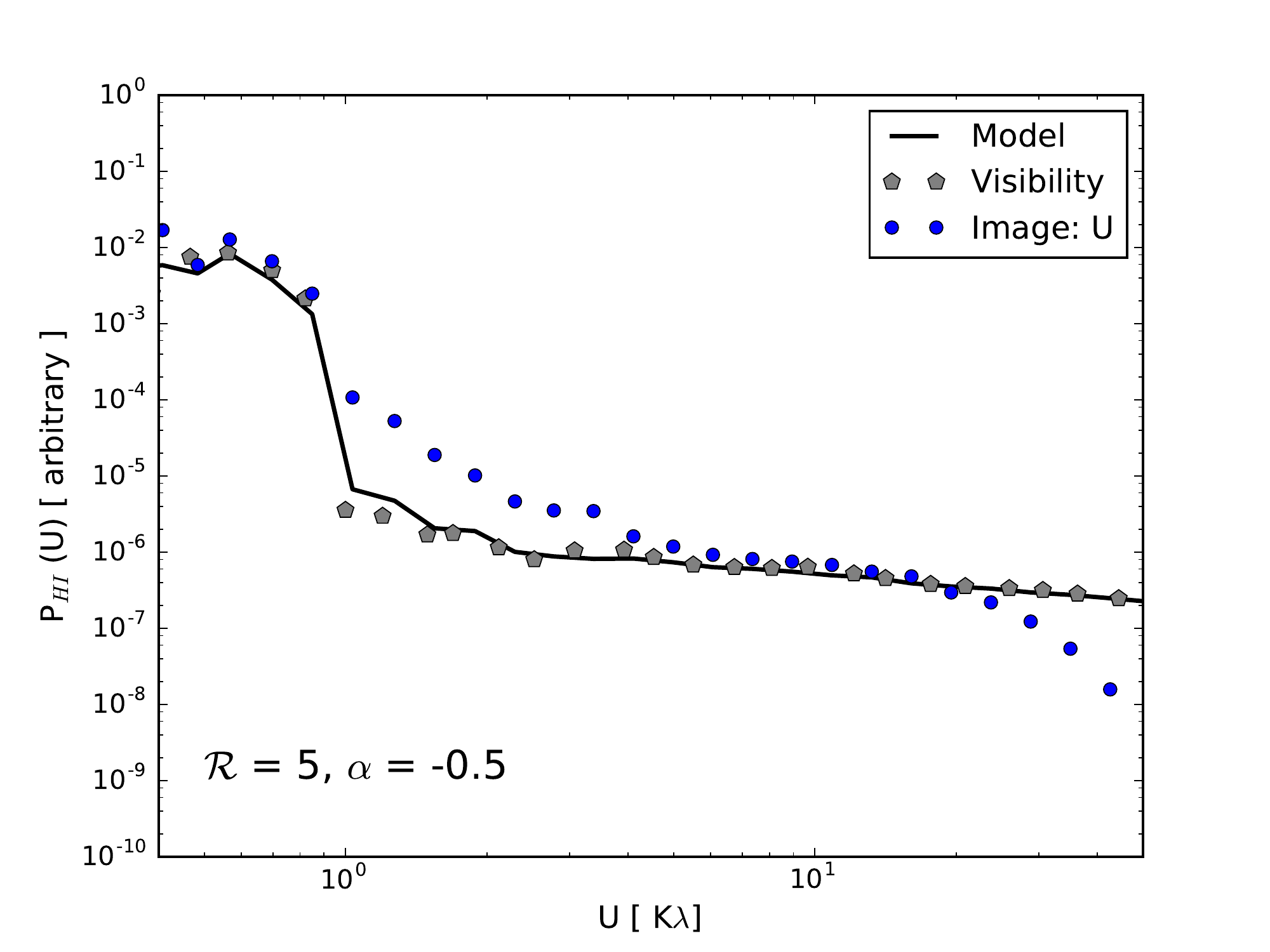}
\includegraphics[scale=.4, trim={0.cm 0. 1.7cm 0.}, clip]{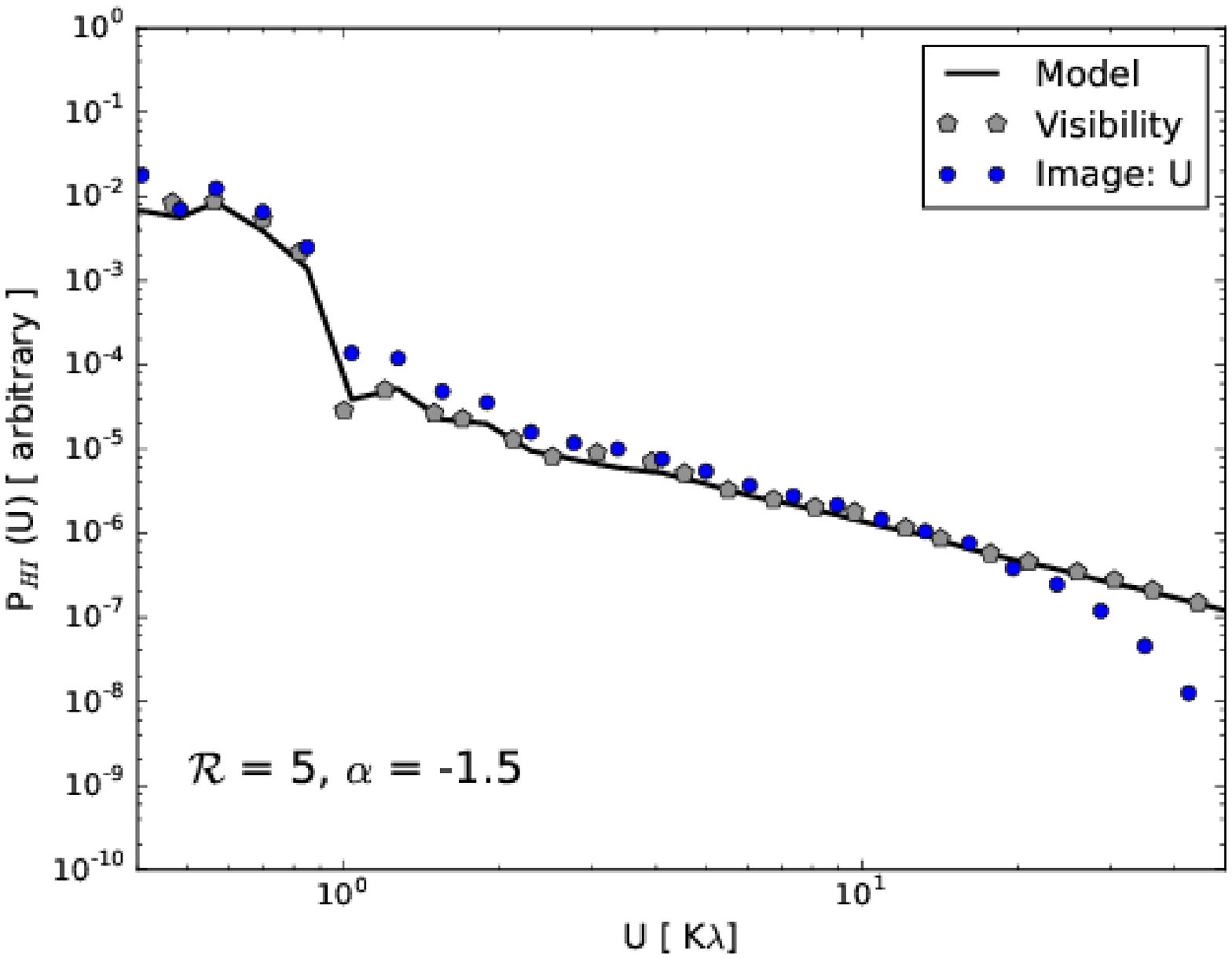}
\includegraphics[scale=.4, trim={0.5cm 0. 2cm 0.}, clip]{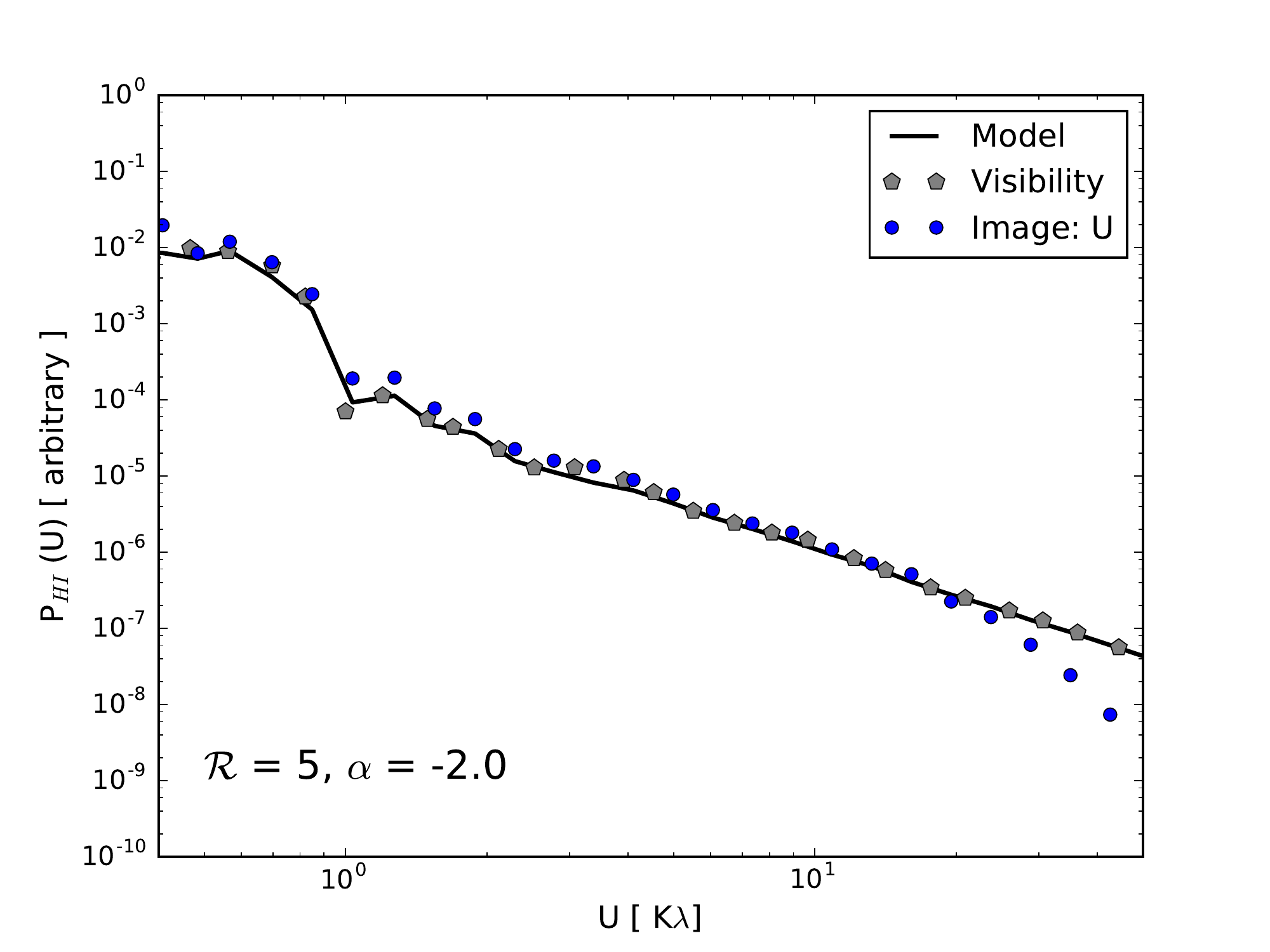}
\includegraphics[scale=.4, trim={0.2cm 0. 1.7cm 0.}, clip]{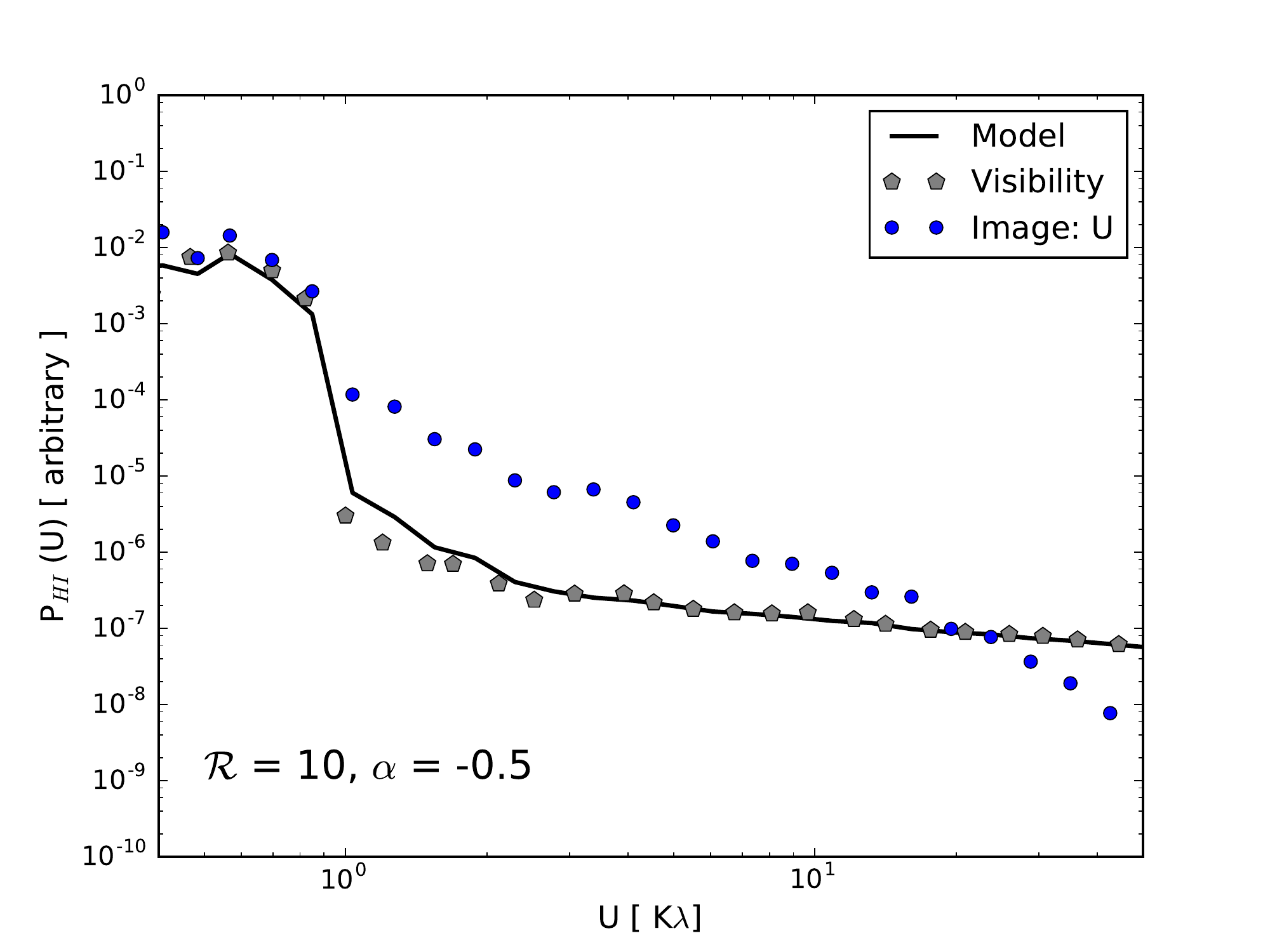}
\includegraphics[scale=.4, trim={0.cm 0. 1.7cm 0.}, clip]{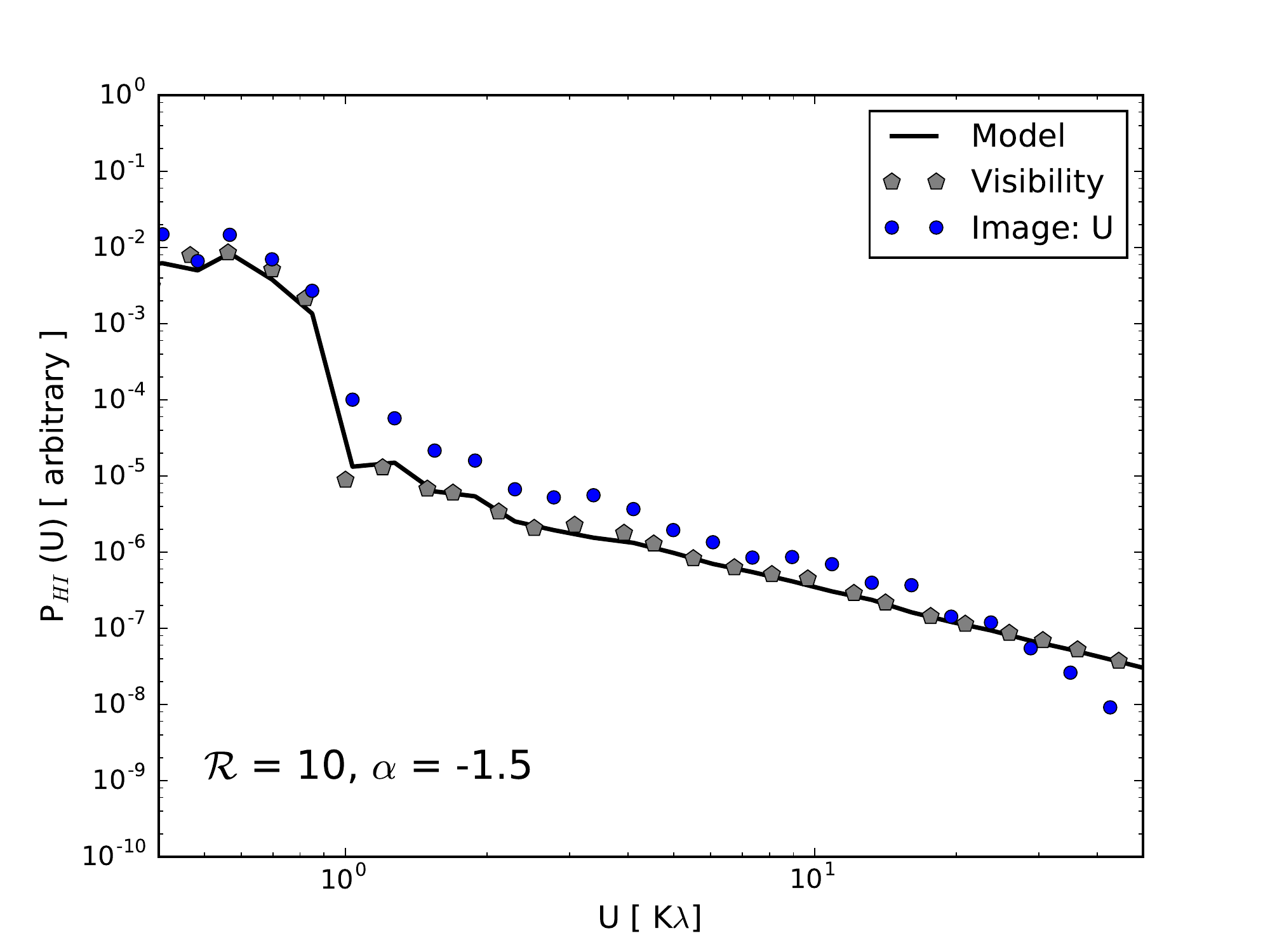}
\includegraphics[scale=.4, trim={0.5cm 0. 2cm 0.}, clip]{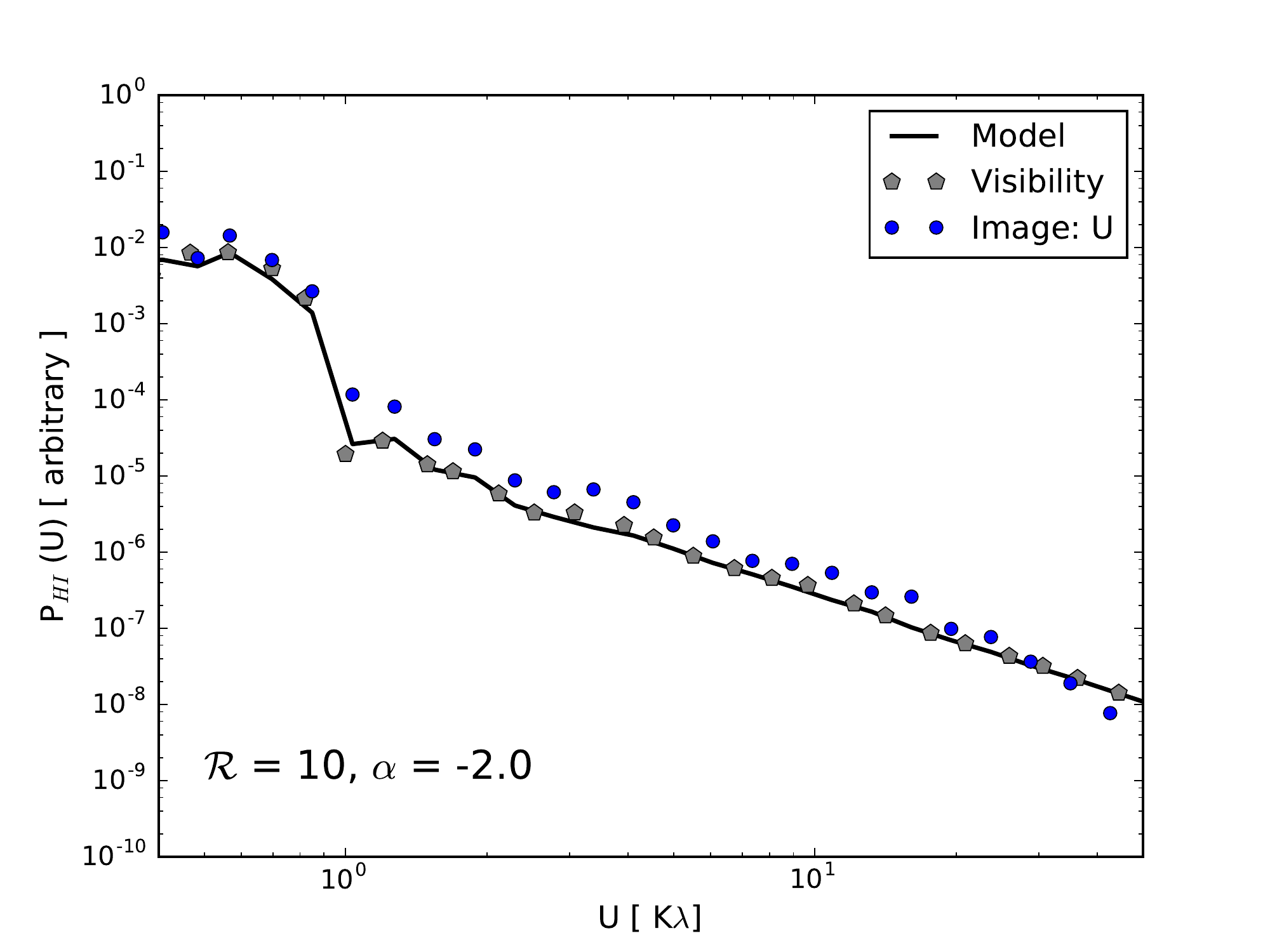}
\caption{Comparison of the power spectrum estimated  from the input model (black solid line), using the visibility based estimator (grey pentagons) and  image based estimators from the uniform weighted image (blue circles) are shown for all six simulations. The amplitudes of the y-axes for different panels are scales arbitrarily to keep the power spectra in the same range in all plots.}
\label{fig:PS6}
\end{figure*}
 
We use the image based power spectrum estimator for all the three reconstructed images from the three weighting schemes.  We correct each of these spectra for the effect of tapering by multiplying them with $1/T(U)^{2}$. We plot in the same figure with circular, square and triangular markers representing the uniform, robust and natural weightings respectively with the corresponding error bands.  At longer baselines, the power falls drastically. This is an effect of the convolution of the CLEAN components with the restoring beam which produces correlation at the pixels at a scale smaller than the beam scale.  The image based estimate with natural weighting scheme is drastically different from the model. The power spectrum estimated using the reconstructed image with the uniform weighting scheme almost follow the model power spectra within the error bars in the baseline range of $1-20\ k\lambda$ and that with the robust weighting scheme is slightly different from the model. As discussed before, we expect the power spectrum to be a power law. To assess how  good are the image based estimates of the power spectra, we fit (chi-square method) a power law function to these spectra between the baseline range $1 - 20\ k\lambda$ and find the best fit value of the power law slope with error bars.   We find the power law index estimated from the image based power spectra vary as $-1.8 \pm 0.1, -2.0 \pm 0.1$ and $-2.2 \pm 0.1$ for the uniform, robust and natural weighted images respectively. The best fit power law index for the visibility based estimate of the power spectra in the same baseline range is $-1.5 \pm 0.1$. These numbers suggests that more or less all three image based estimators deviates  from the power spectrum of the model sky. However, we must note that the uniform weighting scheme preserves the power spectrum of the model with least bias amongst the image based estimators.

\begin{figure*}
 \includegraphics[scale=.43, clip]{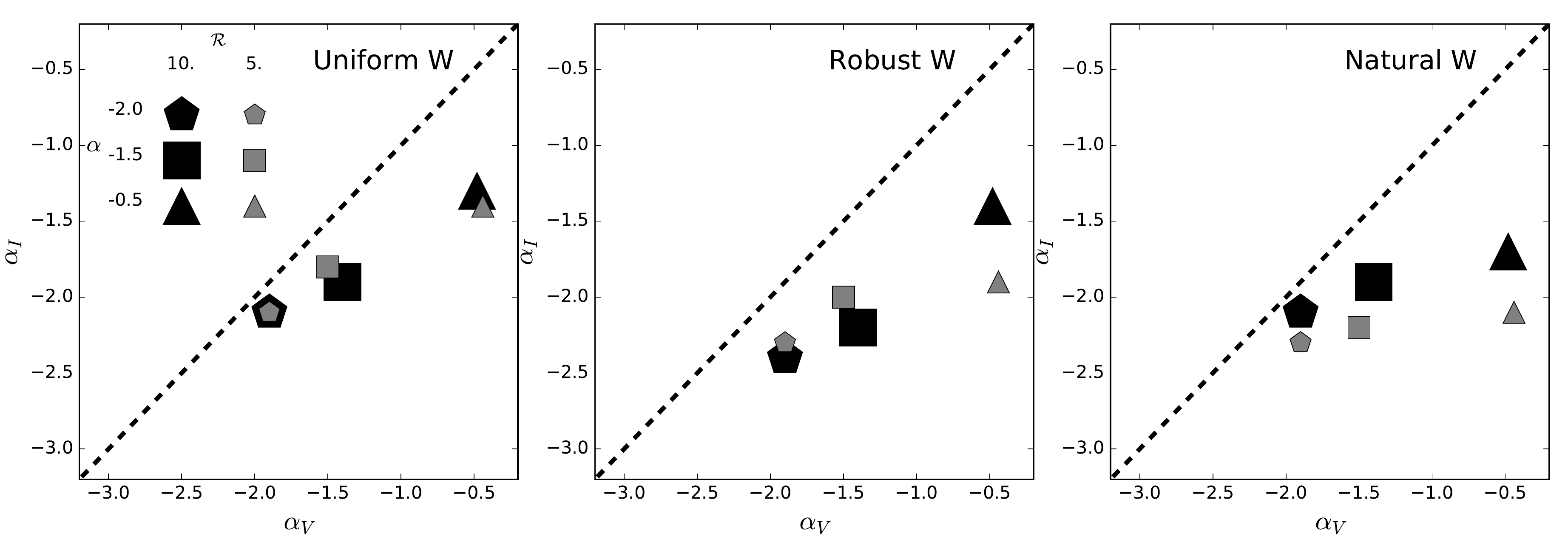}
\caption{The image and visibility based estimates of $\alpha$ are compared for different simulations and weighting schemes. In each panel, we plot the image based estimate $\alpha_{I}$ against the visibility based estimate $\alpha_{V}$ for all six simulations. The dashed line corresponds to an exact match. Meaning of different markers is given in the left most panel. Three panels in this figure correspond to three different weighting schemes, with a left to right giving uniform,  robust and natural weightings respectively.}
\label{fig:pscomp}
\end{figure*}


Figure~(\ref{fig:PS6}) shows the power spectrum estimated from the uniform weighting schemes for all six of our models (blue circles) and from the visibility based estimators (grey pentagons) against the model power spectrum (black solid line). Clearly, the visibility based power spectrum reproduces the model power spectrum almost exactly, whereas the image based estimate of the power spectrum is biased. The bias is visually more prominent for larger values of $\alpha$.  Table~1 gives the values of  the estimated $\alpha$ using visibility and image based estimators for all the different sky models and different weighting. To compare the result from all the six simulations we plot the different estimates of power law slope ($\alpha_{I}$) from the image based estimator against that estimated using the visibility based estimator ($\alpha_{V}$) in figure~(\ref{fig:pscomp}). Three panels in this image correspond to three different weighting schemes. Representations of different markers are given in  the left panel. Clearly, for all models, the uniform weighting scheme performs the best.  We also notice that the $\alpha_{I}$s are systematically smaller than the corresponding $\alpha_{V}$s with  shallower power spectra having a systematically larger bias. 

\section{Discussion and Conclusion}
In this work, we simulate \HI observation of external spiral galaxies to test the efficacy of different estimators that are used to measure the statistical properties of the sky brightness distribution from the radio interferometers. In particular, we have investigated how well we can reconstruct  the large-scale structure of the brightness given by the   window function and the scale dependence of the structures given by the power spectrum of the intensity fluctuations.  In order to estimate the window function, it is essential to reconstruct the sky brightness distribution from the observed visibilities. On the other hand, one can either use the visibilities directly  to estimate the power spectrum, or first estimate the brightness distribution from the visibilities and then use those to estimate the power spectrum. Reconstruction of the brightness distribution is based on several algorithms. We find that with the Cotton-Schwab version of the CLEAN algorithm using natural weighting scheme, the window function is reproduced without any  bias. The visibility based estimator of the power spectrum  reproduces the model power spectrum without any bias. Amongst the image based estimators, the reconstructed image with uniform weighting scheme performs best, however, a general scale dependent bias is observed for all the image based estimators. 

It is clear that for an ideal interferometer with visibilities measured over the entire baseline-plane, that is with complete uv-coverage, the sky brightness distribution can be estimated without any bias. In such a case the image based power spectrum estimators are expected to produce an unbiased result.  Apparent reason for the bias  in the  image based estimates of the power spectrum from a realistic interferometer can be  the incompleteness and non uniformity of the uv-coverage of the interferometer.  Different weighting schemes try to address these issues, however, as our result suggests, for the baseline coverage of the simulated visibilities used here, the weighting methods fail to reproduce the power spectrum of the model sky. Interestingly, the natural weighting gives the best estimate of the window function, while the power spectrum estimated from the image with uniform weighting gives the best approximation to the power spectrum. This can be understood in the following way. The natural weighting gives same weight to all the measured visibilities  in a grid. Since the baseline coverage of the interferometer is more complete at the shorter baselines and falls as the baseline increases, effectively natural weighting produces a larger synthesized beam. Hence, it is expected that the natural weighting would produce a better approximation to the large-scale distribution, while it models the small scale distribution poorly. The effect then is also a redistribution of the  flux across different angular scales. Thereby, the power spectrum at the large-scales, that is at small baselines, gets enhanced, whereas the power spectrum at large baselines is reduced. Effectively, the bias parameter $P_{\mathcal{B}}(\vec{U})$, as discussed in eqn~(\ref{eq:imps}), assumes positive values at smaller baselines and is negative at larger baselines. Uniform weighting, on the other hand, produces smaller restoring beam and hence reproduces relatively more power at the longer baselines compared to the natural weighting. This is observed in the power spectrum estimates.
\begin{figure}
 \includegraphics[scale=.46, clip]{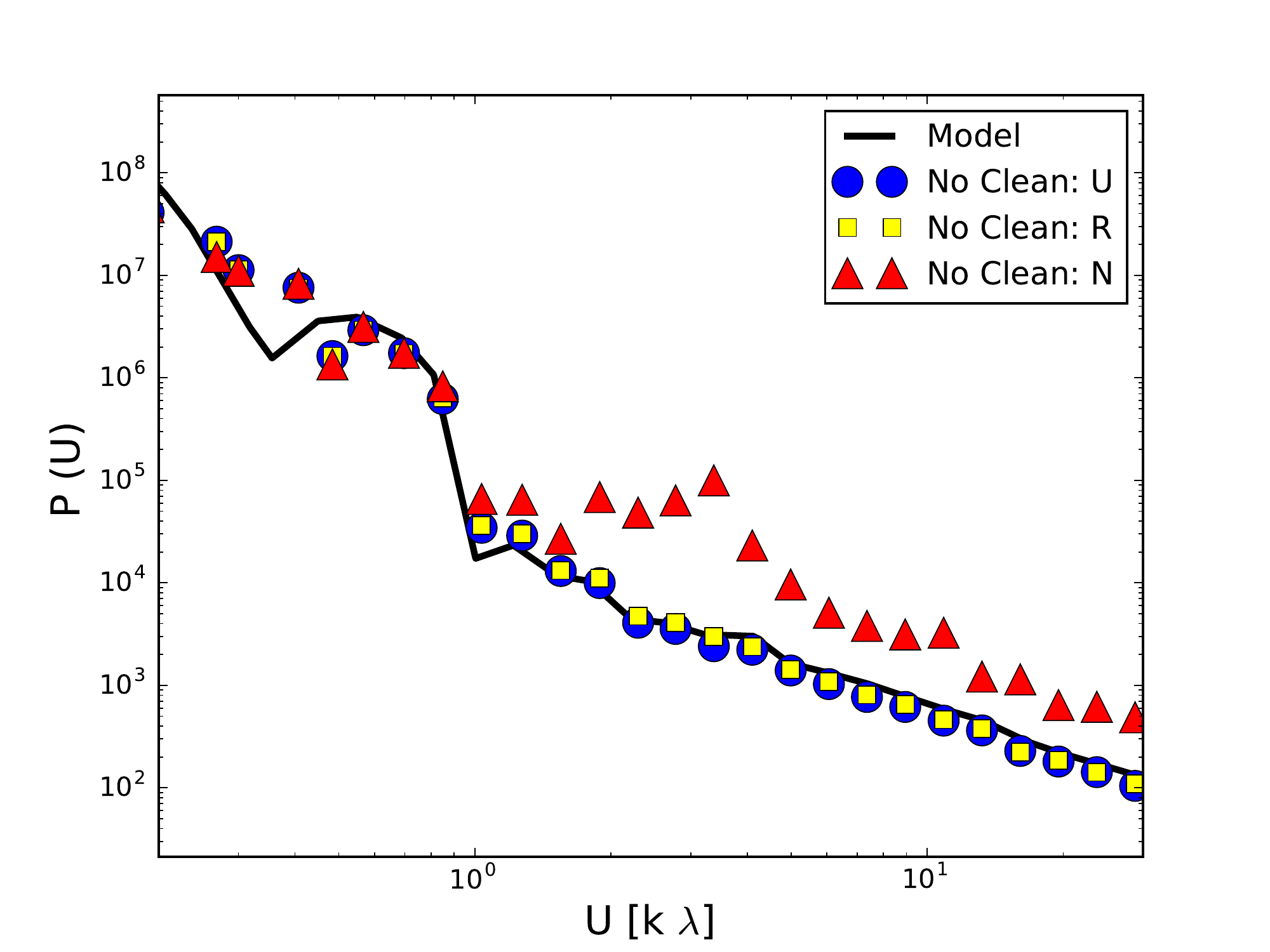}
\caption{Figure showing the beam normalized power spectra from the dirty images with three different  weighting schemes.}
\label{fig:psnc}
\end{figure}

\begin{figure*}
\begin{center}
\includegraphics[scale=.32, clip]{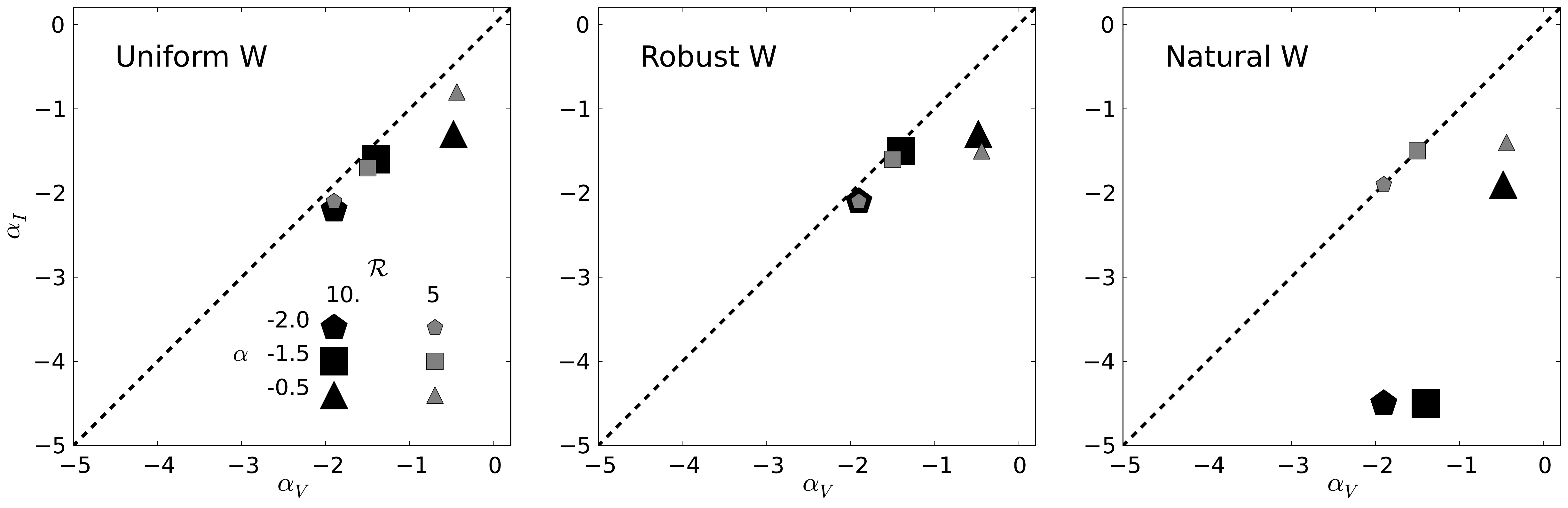}
\caption{The dirty image and visibility based estimates of $\alpha$ are compared for different estimators and weighting schemes. In each panel, we plot  the power law index  $\alpha_{I}$ as estimated from the dirty images against the visibility based estimate $\alpha_{V}$. The dashed line corresponds to an exact match. Meaning of different markers is given in the leftmost panel. Three panels in this figure correspond to three different weighting schemes, with a left to right giving uniform,  robust and natural weightings respectively.}
\label{fig:pscompN}
\end{center}
\end{figure*}

Apparently, we may use the dirty image  as given in eqn.~(\ref{eq:dbeam}) with the image based power spectrum estimator and correct for the effect of $B_{D}(\vec{\theta})$. We approached to do it by first estimating the azimuthally averaged power spectrum of the dirty image using the image based estimator and then dividing that  by the azimuthally averaged  power spectrum of the dirty beam corresponding to the image. Figure~\ref{fig:psnc} shows the power spectra estimated for the simulation with parameters $\mathcal{R} = 5$ and $\alpha  = -1.5$. Clearly, the image with natural weighting does not reproduce the model power spectrum, however, the power spectra calculated from the dirty images with uniform as well as the robust weighting closely resembles the model spectra. To access the robustness of the beam normalized power spectrum estimated this way, we estimated the power law index of the reconstructed  power spectra for all the six simulations and the results are tabulated in Table~(2).  Figure~\ref{fig:pscompN} is a visual representation of these results. It can be clearly seen that as compared to  figure~\ref{fig:pscomp}, which consolidate the results with the cleaned images, the beam normalized power spectra estimated from the dirty images systematically better reproduces the model spectra. Just as before, the uniform weighting gives the best reproduction of the model spectra. However, for larger $\mathcal{R}$ and higher $\alpha$, the uniform weighting estimates are also not much reliable (see Table~2 and figure~\ref{fig:pscompN}). We conclude, though in certain cases, the beam normalized power spectra from the dirty images with uniform weighting would unbiasedly estimate the power spectra of the specific intensity, without prior knowledge of the source brightness distribution, it is better to avoid using such estimators in general.

\begin{table*}
\centering
\begin{tabular}{|l|c|c|c|c|c|c|c|c|c|c|}
\hline
\multicolumn{2}{|l|}{}                                                                       & \multicolumn{3}{c|}{$\alpha = -0.5$}        & \multicolumn{3}{c|}{$\alpha = -1.5$}       & \multicolumn{3}{c|}{$\alpha = -2.0$}       \\ \hline
\multicolumn{2}{|l|}{ $\mathcal{R} $ }    &  U            & R            & N            & U            & R            & N            & U            & R            & N             \\ \hline
\multicolumn{1}{|c|}{\multirow{1}{*}{5}}  &                            & $-0.8\pm0.1$ & $-1.5\pm0.3$ & $-1.4\pm0.7$ & $-1.7\pm0.1$ & $-1.6\pm0.1$ & $-1.5\pm0.3$ & $-2.1\pm0.1$ & $-2.1\pm0.1$ & $-1.9\pm0.2$                                   \\ \hline
\multirow{1}{*}{10}                       &    & $-1.3\pm0.7$ & $-1.3\pm0.7$ & $-1.9\pm0.3$ & $-1.6\pm0.2$ & $-1.5\pm0.4$ & $-4.5^*\pm0.7$ & $-2.2\pm0.1$ & $-2.1\pm0.3$ & $-4.5^*\pm0.7$ \\ \hline
\end{tabular}
\caption{ Table summarising the result of  power spectra estimated using the dirty images.  The header `U', `R' and `N' corresponds to the uniform, robust and natural weighting schemes respectively. Stared values represent that fitting is not good in respective cases.}
\label{tab:fitp2}
\end{table*}

 In case of large bandwidth observations with smooth specific intensity  across frequencies, a method termed as multifrequency synthesis can be effectively applied that drastically improves the effective baseline coverage and hence reduces the CLEAN artefacts. In this work, we model our sky brightness distribution based on \HI emission from external spiral galaxies. In such a case, the specific intensity has structures as a function of  frequency and hence the method of multifrequency synthesis cannot be used. In continuum observations with large bandwidth, however, multifrequency synthesis \citep{1999ASPC..180..419S, 2008ARep...52..951B, 2015AA...581A..59J} and hence, the image based power spectrum may be a useful tool. We also have restricted our analysis for relatively small FOV such that the effect of the w-terms is negligible. For a larger field of view, w-term is known to restrict the baseline range over which the visibility based power spectrum can be used \citep{2010MNRAS.406L..30D}. \citealt{1988AA...200..312W,2008ISTSP...2..793C} have introduced the method of multi-scale CLEAN for image reconstruction in case of extended objects. Multiscale CLEAN tries to find the large-scale structures first from the image and then proceed with smaller scales. While this has been successful and has been used in many cases (e.g \citet{2009AJ....137.4718G}), the subjective choice of the scales at which the CLEAN proceeds remains an open question. An advanced version of multi-scale CLEAN has been reported in \citet{2016Ap&SS.361..153Z}, where adaptive methods are used in choosing the scales and loop gains. We note that this choice of scales may introduce artificial scale dependence in the power spectrum. Hence, we have not considered these algorithms for power spectrum analysis. 

\begin{figure*}
\includegraphics[scale=.42, trim={0.6cm 0. 1.5cm 0.}, clip]{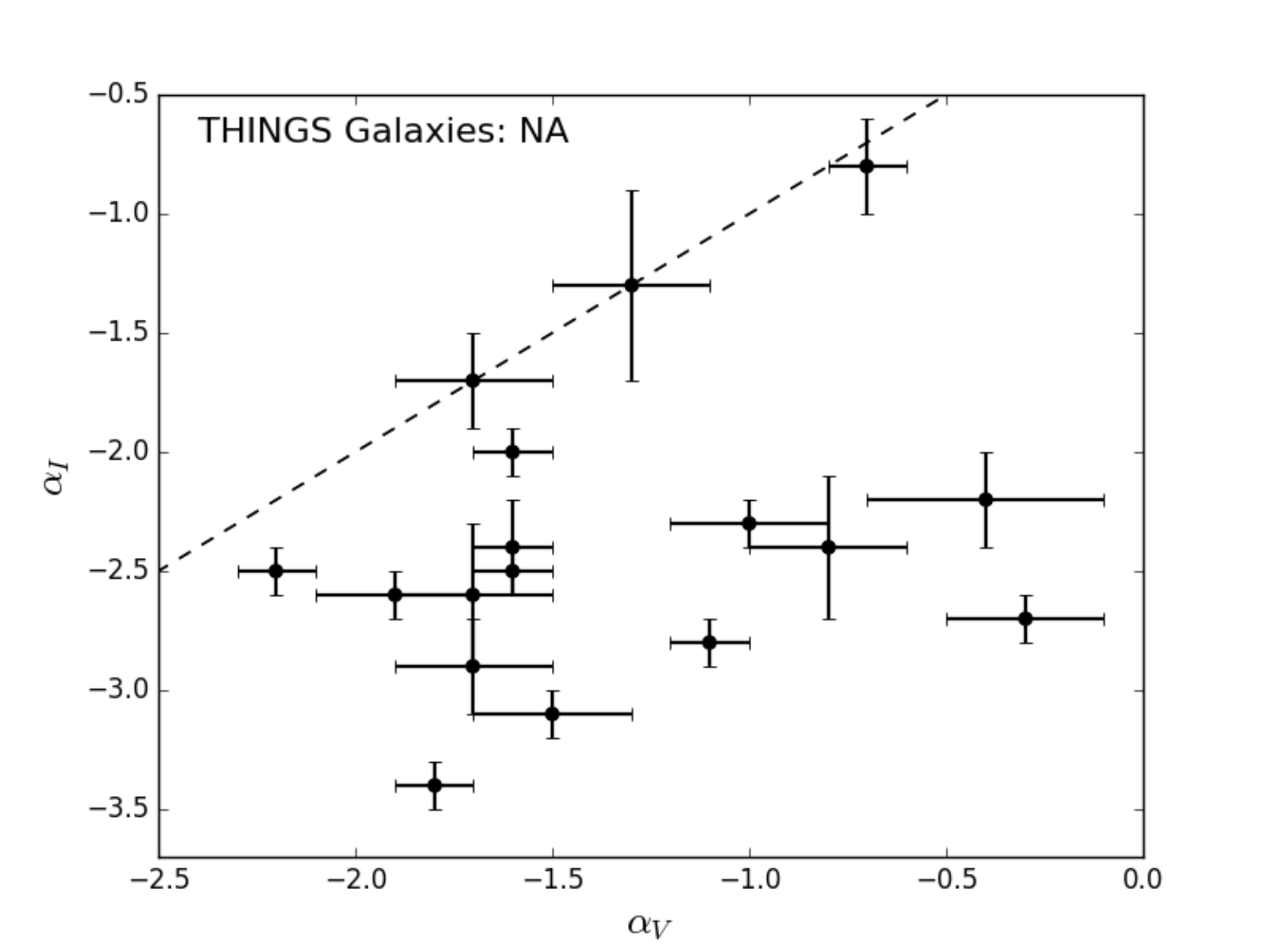}
\includegraphics[scale=.42, trim={0.cm 0. 1.6cm 0.}, clip]{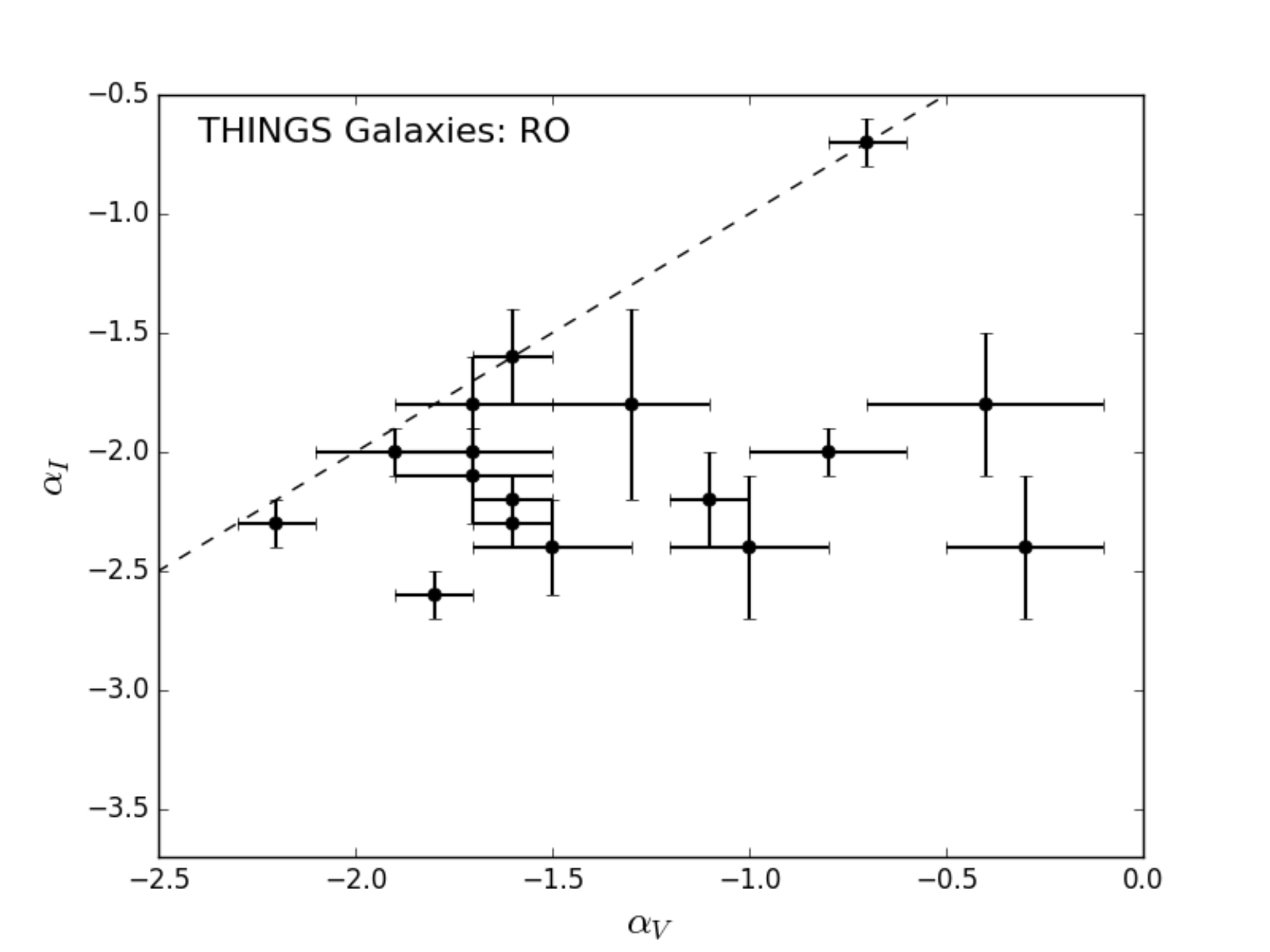}
\caption{Comparison of the power law slope calculated from the visibility to the image based power spectrum for 18 galaxies in the THINGS sample using natural weighted (NA in left) and robust weighted (RO in right) images from the THINGS data archive. Dashed lines correspond to an exact match between the slopes. }
\label{fig:ANA}
\end{figure*}

\citet{2008AJ....136.2563W} have performed a  survey of \HI in a sample of 34 external galaxies using B, C and D configurations of VLA. To reconstruct the images they use the CLEAN algorithm with the natural and robust (with the ROBUST parameter set to 0.5) weighting schemes. The effective resolution of their reconstructed image with the robust weighting scheme is $\sim 6''$. \citet{2013NewA...19...89D} have estimated the slope of the power spectrum ($\alpha_{V}$) of the \HI intensity fluctuation of 18 spiral galaxies from THINGS sample using the same visibility based estimator as we have used here. As our simulation suggests that the visibility based estimator estimates the true power spectrum of the specific intensity fluctuations, we use the measured  $\alpha_{V}$ as a proxy for the value of $\alpha$ for these galaxies. We used the publicly available natural and robust weighted THINGS moment 0 maps of these 18 spiral galaxies to estimate the power spectrum using the image based estimator we have discussed here. These power spectra were well fit by power laws in a similar range of length scales as in the work of \citet{2013NewA...19...89D} and we estimate the corresponding power law slopes ($\alpha_{I}$). We plot the values of $\alpha_{V}$ and $\alpha_{I}$ along with the error bars in  the left and right panel of figure~(\ref{fig:ANA})   for the natural and robust weighted maps. The dashed line corresponds to   $\alpha_{V} = \alpha_{I}$. As it is clear for majority of the galaxies, the data points lie away from the equality line for both cases of natural and robust weighted maps. The image based estimator systematically produces steeper spectra. As expected from our simulation result,  bias in the robust weighted maps are lower but still significant. Moreover, the  figure~\ref{fig:ANA}  compare quite well with figure~\ref{fig:pscomp}, where a similar plot is made for the results of our simulation. As the THINGS archive does not provide any reconstructed moment 0 map estimated using the uniform weighting scheme, we do not show them here. Nevertheless, it is clear from our analysis that in general there exists bias in the image based estimates of the power spectrum. Several authors, including but not limited to \citet{2012ApJ...754...29Z, 2014MNRAS.441..525W, 2017MNRAS.466.1093G} have used the image based estimators to infer about the power spectrum of the sky brightness distribution and observe discrepancy with the visibility based estimators. We believe our investigation answers the reason for these discrepancies.

\begin{figure}
\includegraphics[scale=.46, trim={0.5cm 0. 2cm 0.}, clip]{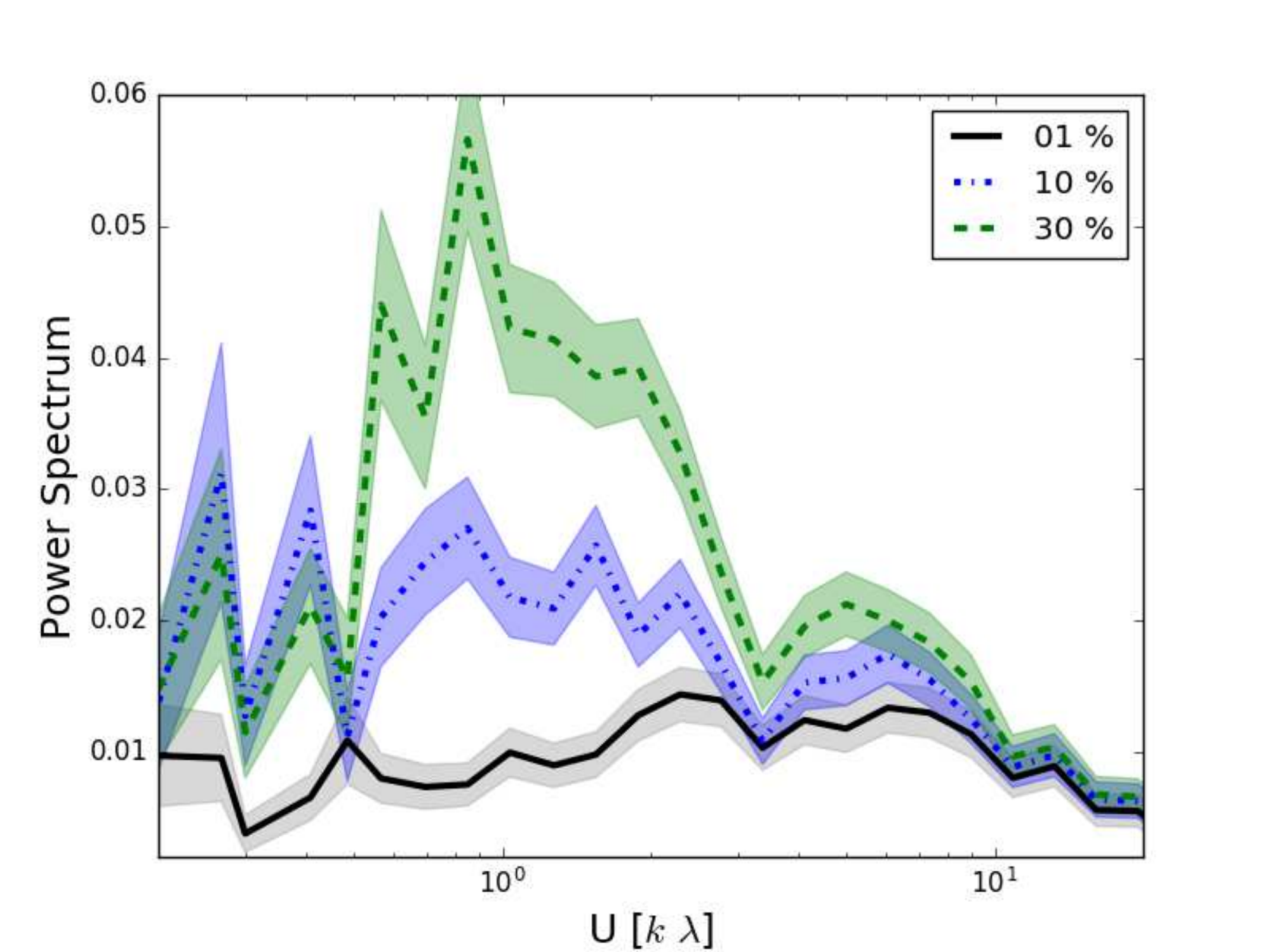}
\caption{The Figure shows the power spectra calculated from the residual images after the point source subtraction for the point sources influencing $1\%$, $10\%$ and $30\%$ of the field of view.}
\label{fig:psinv}
\end{figure}
Apart from the baseline coverage of the interferometer, efficacy of image reconstruction may also depend on the structure of the sky brightness distribution itself. In the CLEAN algorithm, the sky is modelled as a collection of point sources. If the observed sky is a set of isolated unresolved sources, then the visibility function is smooth across baselines. In such cases,  CLEAN is supposed to give an unbiased estimate of the sky. On the other hand, for diffuse emission, the visibility function is expected to be patchy. Observations with  inadequate baseline coverage will lack the full information to model the sky. To test, how much of the sky needs to be filled by sources to see the effect of the baseline coverage,  we model the sky with a collection of point sources uniformly distributed in the field of view. The amplitude of these sources are varied randomly within a decade of flux density,  the absolute flux scale is of no importance here. If we keep the number of point sources small, then they are expected to be isolated and the CLEAN must work well. On the other hand, if we increase the number of point sources, it would start to simulate a diffuse emission and CLEAN may fail to reproduce  an unbiased estimate of the sky. Considering a Gaussian PSF of the interferometer, we assume that each point source influences the nearby pixels within a circle of diameter equal to  $2.5$ times the full-width at half maxima of the PSF. How much of the field of view is covered by the point sources in this way  gives a measure of how diffuse the emission is.  We generate three model sky intensity distributions with $1\%$, $10\%$ and $30\%$ of the sky influenced by the point sources (using robust weighting with a ROBUST=0). Using these model images we simulate the visibilities and reconstruct images keeping the same baseline coverage as  our previous simulations. We subtract the CLEAN component model of the point sources from the reconstructed image to get the residual maps. Power spectra of the residual maps for the three cases are shown in  figure~\ref{fig:psinv}. The shaded region in cases show the corresponding errors. A flat power spectrum is expected if the residual image does not have any correlated noise. We see a significant systematic increase in the amplitude of the residual power spectrum with the number of sources. This demonstrates the limitation of image reconstruction from interferometric data with incomplete uv coverage in reproducing the  structures of the diffuse sky.

As the radio astronomy community plans for larger interferometers, the visibility data volume is expected to grow large. One way of reducing the problem with large data volume is to perform online reconstruction of the images with the instantaneously available visibilities.  Our investigation highlights the problems that may arise in a proper reconstruction of the sky statistics and emphasizes the need for recording the visibilities directly. Visibility based  power spectrum as discussed in this work has limited use when power spectrum of a selective part of the telescope field of view is required to be estimated. \citet{2016MNRAS.463.4093C} has developed a visibility based Tapered Gridded Estimator (TGE) that uses a tapering window to reduce the response of the sky outside it. We have worked on the modifications of TGE  limiting the tapering function to the required part of the field of view of interest and selectively estimate the power spectrum using the visibility based method and the results will be reported in a separate paper.

\section*{Acknowledgement}
PD would like to acknowledge numerous useful discussion with Wasim Raja and Aritra Basu. PD will also like to acknowledge Apurba Bera, Barnali Das, Saheb Ghosh and Surajit Mandal for providing him needful hospitality to finish writing the paper. PD acknowledge the support of DST-INSPIRE faculty fellowship from DST, India for this work. MN acknowledge DST-INSPIRE fellowship for funding this work.


\end{document}